\documentclass[a4paper,11pt]{article}
\usepackage{jheppub} 
\usepackage{lineno}

\title{\boldmath Many-body chaos and pole-skipping in holographic charged rotating fluids}

\author[a,*]{Hong-Da Lyu,}
\author[b,*]{Jun-Kun Zhao,}
\author[b,c,d]{Li Li}
\affiliation[a]{Key Laboratory of Particle Physics and Particle Irradiation (MOE),
Institute of Frontier and Interdisciplinary Science,
Shandong University, Qingdao, Shandong 266237, China}
\affiliation[b]{Institute of Theoretical Physics, Chinese Academy of Sciences, Beijing 100190, China}
\affiliation[c]{School of Fundamental Physics and Mathematical Sciences, Hangzhou Institute for Advanced Study, UCAS, Hangzhou 310024, China}
\affiliation[d]{School of Physical Sciences, University of Chinese Academy of Sciences, No.19A Yuquan Road, Beijing 100049, China}

\affiliation[*]{These authors contributed equally to this work.}

\emailAdd{hongdalyu@sdu.edu.cn,junkunzhao@itp.ac.cn,liliphy@itp.ac.cn}

\abstract{Recent developments identify pole-skipping as a `smoking-gun' signature of the hydrodynamic nature of chaos, offering an alternative way to probe quantum chaos in addition to the out-of-time-ordered correlator (OTOC). We study the quantum chaos and pole-skipping phenomenon in the strongly coupled charged rotating fluids, holographically dual to rotating black holes with nontrivial gauge field. We find that the near-horizon equation governing energy-density fluctuations differs from the source-less shock wave equation determining the OTOC, which depends on the $U(1)$ gauge choice. This discrepancy is eliminated under an appropriate boundary condition on the $U(1)$ gauge potential at the event horizon, as required by the vanishing of Wilson loop at the Euclidean horizon. We further investigate the dependence of the butterfly velocity on the charge and rotation parameters in a specific black hole configuration--the Cveti\v{c}-L\"{u}-Pope solution.}

\begin{document}
\maketitle
\flushbottom

\section{Introduction}

Chaos, the butterfly effect, plays a pivotal role in understanding thermalization and information scrambling in finite-temperature many-body systems. One of the most important probes of quantum chaos is the out-of-time-order correlator (OTOC), which for a broad class of systems exhibits exponential growth~\cite{Larkin:1969,Shenker:2013pqa,Roberts:2014isa,Shenker:2014cwa}
\begin{equation} \label{eq:otoc}
\langle V(t,\vec{x}) W(0) V(t,\vec{x}) W(0) \rangle_\beta \sim 1 - \epsilon e^{\lambda_L \left( t - |\vec{x}|/v_B \right)} \,,
\end{equation}
where $V(t,\vec{x})$ and $W(t,\vec{x})$ are generic local operators and the correlator is computed in a thermal ensemble with temperature $T=1/\beta$. Note that $\epsilon$ is a small parameter that scales inversely with the number of degrees of freedom of the theory. The OTOC given in~\eqref{eq:otoc} defines two quantities: the Lyapunov exponent $\lambda_L$, which characterizes the exponential growth of chaos in time, and the butterfly velocity $v_B$, which measures the information spread in space. For strongly coupled quantum systems admitting holographic duals, quantum chaos can be rigorously analyzed through the Dray-'t Hooft shock wave formalism~\cite{Dray:1984ha,Sfetsos:1994xa}. Remarkably, gravitational calculations in such systems reveal the maximal Lyapunov exponent $\lambda_L = 2\pi T$~\cite{Shenker:2013pqa,Roberts:2014isa}, saturating the chaos bound~\cite{Maldacena:2015waa}. Various systems saturating the chaos bound have been identified, including: holographic systems~\cite{Shenker:2013pqa,Jensen:2016pah,Maldacena:2016upp}, the Sachdev-Ye-Kitaev model and its cousins~\cite{Maldacena:2016hyu,Kitaev:2017awl,Gu:2016oyy,Davison:2016ngz}, conformal field theories~\cite{Turiaci:2016cvo,Haehl:2018izb,Haehl:2019eae}, etc. Nevertheless, the distinctive features and universal characteristics of these maximally chaotic systems remain under active investigation.

Remarkably, a profound connection between late-time hydrodynamics and early-time quantum chaos emerges through the \emph{pole-skipping phenomenon}~\cite{Grozdanov:2017ajz,Blake:2017ris,Blake:2018leo}. At specific point in the complex momentum space, the Lyapunov exponent $\lambda_L$ and butterfly velocity $v_B$ can be directly identified, as first observed numerically in Schwarzschild AdS$_5$ black hole~\cite{Grozdanov:2017ajz} and then extended to Einstein gravity coupled to matter fields~\cite{Blake:2018leo}. Express the retarded energy density two-point Green function as 
\begin{equation} \label{eq:green}
G^R(\omega, k) = \frac{b(\omega,k)}{a(\omega,k)} \,,
\end{equation}
where the poles of the Green functions $G^R$ corresponds to $a(\omega, k)=0$. However, at the pole-skipping point
\begin{equation} \label{eq:greenps}
\omega_* = i\lambda_L \,, \quad k_* = \frac{i \lambda_L}{v_B} \,,
\end{equation}
both $a(\omega_*,k_*)=b(\omega_*, k_*)=0$, the would-be pole becomes  skipped due to the multiplication of a zero, leading to a non-unique retarded Green function~\cite{Blake:2018leo}. Moreover, the phenomenon of pole-skipping offers a natural explanation for the relationship between butterfly velocity and hydrodynamic diffusivity observed in earlier studies~\cite{Blake:2016wvh,Blake:2016jnn,Blake:2017qgd} (see also~\cite{Gu:2016oyy,Davison:2016ngz}). It has also been shown to impose constraints on transport bounds~\cite{Grozdanov:2020koi}. Since then, various aspects of pole-skipping have been extensively investigated in holographic systems~\cite{Grozdanov:2018kkt,Grozdanov:2019uhi,Blake:2019otz,Natsuume:2019sfp,Ahn:2019rnq,Wu:2019esr,Abbasi:2019rhy,Abbasi:2020ykq,Abbasi:2020xli,Jansen:2020hfd,Jeong:2021zhz,Ageev:2021xjk,Wang:2022xoc,Yuan:2024utc,Pant:2025zhw,Ahn:2025exp} and boundary field theories~\cite{Haehl:2018izb,Jensen:2019cmr,Das:2019tga,Haehl:2019eae,Ramirez:2020qer,Choi:2020tdj,Asplund:2025nkw}, reinforcing the proposal to use the hydrodynamic nature of chaos as a defining characteristic of maximally chaotic systems~\cite{Blake:2017ris,Blake:2021wqj,Knysh:2024asf}.

In holographic theories, the pole-skipping phenomenon stems from the decoupling of energy density perturbations from other modes at the event horizon precisely at $\omega = i\lambda_L$, yielding equations identical to the source-less shock wave equation~\cite{Blake:2018leo,Wang:2022mcq}. It should be noted, however, that the analysis has focused primarily on static spacetimes in Einstein-Maxwell-Scalar theory. Moreover, the statement in~\cite{Blake:2018leo} relies on the condition $T_{vr} h_{vv}-\delta T_{vv}=0$, where $T_{vr}$ and $\delta T_{vv}$ denotes components of energy-momentum tensor $T_{ab}$ of matter fields and its perturbation, while $h_{vv}$ corresponds to the $vv$ component of metric perturbations in ingoing Eddington-Finkelstein coordinates. This condition remains an assumption that has not been rigorously justified for generic backgrounds, raising the possibility that pole-skipping may not hold universally in maximally chaotic systems. On the other hand, rotation represents a fundamental and ubiquitous feature of physical systems, spanning scales from atomic nuclei to galaxies. Even strongly coupled systems such as quark-gluon plasmas can exhibit vortical fluid behavior~\cite{STAR:2017ckg,Liang:2004ph}. This broad relevance has motivated substantial efforts to incorporate angular momentum into hydrodynamic and holographic models~\cite{Florkowski:2017ruc,Florkowski:2018fap,Hawking:1998kw,Garbiso:2020puw}. 

Unlike static configurations, rotation modifies the background geometry in a nontrivial way: the metric acquires non-diagonal components, which significantly complicates the analysis of perturbations. In such cases, angular momentum and its conjugate variable, the angular velocity, enter the first law of black hole thermodynamics. Furthermore, rotation introduces anisotropy into the system, resulting in two distinct pressures--a property analogous to that observed in anisotropic black branes~\cite{Critelli:2014kra,Giataganas:2017koz,Cai:2024tyv}. Given the importance of rotating black holes, the pole-skipping phenomenon has been studied in several notable geometries, including the BTZ~\cite{Liu:2020yaf,Mezei:2019dfv,Jahnke:2019gxr}, Kerr-AdS$_4$~\cite{Blake:2021hjj}, and Myers–Perry-AdS$_5$ black holes~\cite{Amano:2022mlu}. However, these are all vacuum solutions and therefore lack matter fields contribution\footnote{Studies of pole-skipping in charged magnetic black holes with vanishing angular velocity can be found in~\cite{Abbasi:2019rhy,Abbasi:2023myj,Zhao:2025gej}.}. As a result, the influence of matter energy-momentum tensor on pole-skipping phenomenon in rotating background geometries remains unexplored. This motivates our study on the relation between pole-skipping and quantum chaos in rotating spacetime, particularly with the presence of matter fields.

The most straightforward example of a rotating black hole with matter fields is the Kerr-Newman (KN) solution in the four-dimensional Einstein-Maxwell theory~\cite{Carter:1968ks,Hawking:1998kw,Caldarelli:1999xj}. However, its metric functions depend on both radial and angular coordinates, which complicates the analysis. Especially, similar to the Kerr AdS$_4$ case~\cite{Blake:2021hjj}, the analytical treatment for KN black hole may be restricted to the small angular momentum regime. In contrast, five-dimensional rotating black holes with equal angular momenta exhibit enhanced symmetry, reducing the background to cohomogeneity-1 and enabling a more simpler treatment~\cite{Emparan:2008eg}. Although closed-form five-dimensional Kerr-Newman solutions remain elusive, exact rotating black hole~\emph{i.e.} Cvetič-Lü-Pope (CLP) solutions become readily accessible through the inclusion of Chern-Simons term~\cite{Cvetic:2004hs,Madden:2004ym}. In the present work, we focus on the five-dimensional rotating black holes with equal angular momenta in Einstein-Maxwell-Chern-Simons (EMCS) theory. Crucially, our analysis holds universally—independent of the specific value of the Chern-Simons coupling. We derive the horizon equation of energy density fluctuations and the shock wave equation. Interestingly, we find the decoupled horizon equation of energy density fluctuations obtained via pole-skipping analysis differs from the source-less shock wave equation, which can depend on the $U(1)$ gauge freedom of the matter field. This issue is fixed by considering a physical boundary condition for the gauge potential at the black hole event horizon through the Wilson loop argument, leading to a perfect agreement between pole-skipping and OTOC. Moreover, we show that finite-size effect results in a complex butterfly velocity, with its imaginary part suppressed in the large black hole limit. We then perform an explicit computation of the butterfly velocities for CLP black holes, examining their dependence on both the charge and rotation parameters.

The paper is organized as follows. In Section~\ref{sec2}, we introduce the five-dimensional charged rotating black hole. Section~\ref{sec:3} is devoted to the standard computation of OTOC and the associated Lyapunov exponent and butterfly velocity. In Section~\ref{sec:4}, we study the pole-skipping phenomena and determinate the butterfly velocity from the pole-skipping point. In Section~\ref{sec:5}, the equivalence between the OTOC and pole-skipping results is demonstrated, highlighting the importance in imposing physical boundary condition on the $U(1)$ gauge potential at the black hole event horizon.  An explicit computation of the butterfly velocity for the CLP black hole is presented in Section~\ref{sec:6}. We conclude and discuss the open questions in Section~\ref{sec:7}. More technical details are provided in Appendices~\ref{app:shockwave} and~\ref{app:clp}.

\section{Model and setup}\label{sec2}

The action of the 5-dimensional ECMS theory reads
\begin{eqnarray}
\begin{split}\label{eq:action}
S=\frac{1}{16\pi G_N} \int & d^5x \sqrt{-g} \Big( R+\frac{12}{L^2}- \frac{1}{4}F_{ab}F^{ab} +\frac{k}{24} \epsilon^{abcde}A_a F_{bc} F_{de} \Big) \,, 
\end{split}
\end{eqnarray}
where $G_N$ is the Newton's constant, $L$ is the AdS radius. Also, the gauge field strength is $F_{ab}=\partial_aA_b -\partial_bA_a$, and $k$ is the Chern-Simons (CS) coupling. For $k=k_{susy}=2/\sqrt{3}$, the above action arises as the bosonic part of five dimensional minimal supergravity~\cite{Buchel:2006gb}. The CS coupling constant $k$ remains a free parameter in our analysis, as the results are independent of its value.

The equations of  motion are given by
\begin{eqnarray} 
\begin{split} \label{eq:eom}
\mathcal{E}_{ab} \equiv R_{ab} +\frac{4}{L^2}g_{ab} -\frac{1}{2} \left(F_{ac}F_b^{\ c}-\frac{1}{6}g_{ab}F^2 \right) &=0  \,,   \\ 
\nabla_{b}F^{ba} +\frac{k}{8} \epsilon^{abcde}F_{bc}F_{de} &=0  \,. 
\end{split} 
\end{eqnarray}
Generically, the five dimensional rotating black holes admit two distinct angular momenta and the isometry group is $\mathbb{R}_t\times U(1)^2$, where $\mathbb{R}_t$ represents the time translation. Constructing rotating black hole solutions in the general case is a challenging problem. However, if the two angular momenta are equal and non-vanishing then the $U(1)^2$ associated with the corresponding 2-planes is enhanced to a non-abelian $U(2)\simeq SU(2)\times U(1)$ symmetry. In such case, the background solutions become co-homogeneity one and depend on the radial coordinates only. The general background ansatz for the metric and gauge field that incorporates a local $SU(2)\times U(1)$ symmetry reads~\cite{Madden:2004ym,Blazquez-Salcedo:2016rkj}
\begin{eqnarray} 
\begin{split} \label{eq:ansatz}
ds^2&=-F_0dt^2 +\frac{dr^2}{F_1} +\frac{F_2}{4} \left(\sigma_1^2+\sigma_2^2\right) +\frac{F_3}{4}\left(\sigma_3-\Omega dt \right)^2 \,, \\
A&=A_t dt +\frac{1}{2} A_\psi \sigma_3 \,, 
\end{split} 
\end{eqnarray}
where $\sigma_1=\cos{\psi} d\theta+\sin\psi \sin\theta d\phi \,,\sigma_2=-\sin{\psi} d\theta+\cos\psi\sin\theta d\phi\,,
\sigma_3=d\psi +\cos\theta d\phi\,$ are left-invariant one-forms on a 3-sphere expressed in terms of the angular coordinates $(\theta,\phi,\psi)$. The range for Euler angles is $0\leq \theta\leq \pi, 0\leq \phi \leq 2\pi$, and $ 0\leq \psi\leq 4\pi$. Note that $F_0, F_1, F_2, F_3, \Omega, A_t$ and $A_\psi$ depend solely on the radial coordinates $r$. The metric ansatz in~\eqref{eq:ansatz} possesses a reparametrization symmetry, which could be used to fix one of the functions $F_i(r) (i=0,1,2,3)$. A standard choice is to set $F_2(r)=r^2$. However, for our purposes, we will not explicitly fix any particular function but instead impose the condition $F_1(r)/F_0(r)=F_3(r)/F_2(r)$. One can recover the familiar unit $3$-sphere coordinates $(\hat{\theta},\hat{\phi},\hat{\psi})$ by defining $\hat{\theta} =\theta/2 \,,\hat{\psi} -\hat{\phi} =\phi \,,\hat{\psi} +\hat{\phi} =\psi$, from which one has
\begin{eqnarray}
\frac{1}{4} \left( \sigma_1^2 +\sigma_2^2 +\sigma_3^2 \right) = d\hat{\theta}^2 +\sin^2\hat{\theta} d\hat{\phi}^2 +\cos^2 \hat{\theta} d\hat{\psi}^2    \,,\qquad  \frac{\sigma_3}{2}=  \sin^2\hat{\theta} d\hat{\phi} +\cos^2\hat{\theta} d\hat{\psi} \,. 
\end{eqnarray}
Then the background~\eqref{eq:ansatz} can be rewritten as
\begin{equation} 
\begin{split} 
ds^2 &= -F_0 dt^2 + \frac{dr^2}{F_1} +F_2 d\hat{\theta}^2 +\left(F_2 -F_3\right) \sin^2\hat{\theta}\cos^2\hat{\theta} \left( d\hat{\psi} -d\hat{\phi} \right)^2 \\
& \qquad +F_3 \sin^2\hat{\theta} \left( d\hat{\psi} - \frac{\Omega}{2} dt \right)^2 +F_3 \cos^2\hat{\theta} \left( d\hat{\phi} -\frac{\Omega}{2} dt \right)^2 \,, \\
A &= A_t dt +A_\psi \left( \sin^2\hat{\theta} d\hat{\phi} +\cos^2\hat{\theta} d\hat{\psi} \right) \,,
\end{split} 
\end{equation}
from which the presence of two equal angular momenta is manifest.

It is worth noting that for $k=0$, \eqref{eq:ansatz} corresponds to the five dimensional Kerr–Newman black hole, for which only numerical solutions with equal-magnitude angular momenta are accessible~\cite{Kunz:2007jq,Blazquez-Salcedo:2016rkj}. Moreover, obtaining analytical black hole solutions for arbitrary $k$ is generally challenging. However, at the supergravity value $k=2/\sqrt{3}$, the equations admit an exact charged rotating solution: the CLP black hole~\cite{Cvetic:2004hs,Madden:2004ym}. When the gauge field is turned off, this rotating black hole~\eqref{eq:ansatz} reduces to the Myers-Perry-AdS$_5$ solution with equal angular momentum~\cite{Amano:2022mlu,Hawking:1998kw}.

As $r\to\infty$, we require the geometry to be asymptotically AdS$_5$. In particular, the metric functions $F_0(r)$ and $F_1(r)$ scale as $r^2/L^2$, while $F_2(r)$ and $F_3(r)$ asymptotically approach $r^2$. Furthermore, we impose the condition $\Omega(r) \to0$ at the AdS boundary, indicating that the solution is expressed in a non-rotating frame at infinity. Thus, the conformal boundary metric takes the form
\begin{equation} \label{eq:sbdy}
ds^2_{bdy} = -dt^2 + \frac{L^2}{4} \left( \sigma_1^2 +\sigma_2^2 +\sigma_3^2 \right)=-dt^2+L^2(d\hat{\theta}^2 +\sin^2\hat{\theta} d\hat{\phi}^2 +\cos^2 \hat{\theta} d\hat{\psi}^2 )  \,,
\end{equation}
corresponding to a dual system on a 3-sphere of radius $L$.

Assume the outer horizon is located at $r_h$, where both $F_0(r_h)=0$ and $F_1(r_h)=0$. Then, the Hawking temperature and the angular velocity of the horizon, measured relative to the non-rotating frame~\eqref{eq:sbdy} at the AdS boundary~\cite{Gibbons:2004ai}, are given by
\begin{eqnarray} \label{eq:tem}
T =\beta^{-1}= \frac{F_0'(r)}{4\pi}\sqrt{\frac{F_1(r)}{F_0(r)}} \Bigg|_{r=r_h} \,, \quad  \Omega_H= \Omega(r_h) \,.
\end{eqnarray}
This rotating black hole~\eqref{eq:ansatz} provides an example whose AdS/CFT dual describes a four dimensional conformal field theory on $R^t\times S^3$ with finite temperature, finite charge density, and equal angular velocities~\cite{Hawking:1998kw}.

\section{Shock wave and OTOC}\label{sec:3}
The butterfly effect in the boundary quantum field theory can be probed holographically through the behavior of infalling particles towards the black hole~\cite{Shenker:2013pqa}. The exponential growth in OTOC after the scrambling corresponds to the late time near horizon boost of the particles which will eventually back-reacte the geometry~\eqref{eq:ansatz}. Then, the chaos parameters in OTOC, Lyapunov exponent $\lambda_L$ and the butterfly velocity $v_B$, can be read off from the form of shock wave geometry. To perform the shock wave analysis, it is convenient to introduce the Kruskal-like coordinates to make the horizon manifestly smooth~\cite{Shenker:2013pqa,Roberts:2014isa,Amano:2022mlu}.

Define the co-rotating angular coordinate $\tilde\psi$ and the associated one form $\tilde\sigma_3$ by
\begin{align}
 \tilde\psi = \psi - \Omega_H t\,, \quad   \tilde\sigma_3 = d\tilde\psi + \cos\theta d\phi =\sigma_3 -\Omega_H dt \,.
\end{align}
In co-rotating coordinates $(t, r, \theta, \phi,\tilde\psi)$, the background metric take the form of
\begin{align}\label{trmetric}
ds^2=-F(r) dt^2 +\frac{dr^2}{F_1(r)} +\frac{F_2(r)}{4}\left( \sigma_1^2+\sigma_2^2\right)+\frac{F_3(r)}{4}\tilde\sigma_3^2 +\frac{\Omega_H- \Omega(r)}{2} F_3(r) dt d\tilde\sigma_3 \,, 
\end{align}
where $F(r)= F_0(r) -\frac{1}{4}F_3(r)\left[\Omega(r) -\Omega_H \right]^2$. In the above co-rotating frame, the black hole horizon corresponds to the Killing horizon generated by the Killing vector $\chi^a=(\frac{\partial}{\partial t})^a$ as visible from~\eqref{trmetric}. Equivalently, in the original coordinates $(t,r, \theta,\phi,\psi)$ of~\eqref{eq:ansatz}, the Killing vector takes the form of $\chi^a=(\frac{\partial}{\partial t})^a + \Omega_H(\frac{\partial}{\partial \psi})^a $. Moreover, the ingoing $v$ and outgoing $u$ Eddington-Finkelstein coordinates are defined as
\begin{align}\label{eq:EF}
dv = dt + dr_\ast \,,\quad  du = dt - dr_\ast \,,
\end{align}
where the tortoise coordinate $dr_* = dr / \sqrt{F(r) F_1(r)}$. Hence, we define the following Kruskal coordinates
\begin{align}
    U= - e^{-2\pi T u}\,,\quad  V= e^{2\pi T v}  \,,
\end{align}
where one has $UV=-e^{4\pi T r_*}$ from~\eqref{eq:EF}. Consequently, the original background ansatz~\eqref{eq:ansatz} in Kruskal-Szekeres coordinates is given by
\begin{eqnarray}
\begin{split}\label{eq:Kruskal}
ds^2_{KS} &=  \frac{F(r)}{4\pi^2T^2 U V} dU dV + B(U,V) \left( \frac{dV}{V} -\frac{dU}{U} \right) \tilde\sigma_3  +\frac{F_2(r)}{4} (\sigma_1^2+ \sigma_2^2) + \frac{F_3(r)}{4} \tilde\sigma_3^2\,, \\
A_{KS} &=\frac{2A_t(r)+\Omega_H A_\psi(r)}{8\pi T}\left(\frac{dV}{V}-\frac{dU}{U}\right) + \frac{A_\psi(r)}{2} \tilde\sigma_3\,,
\end{split}
\end{eqnarray}
where $B(U,V)= F_3(r) \big[\Omega_H -\Omega(r) \big]/(8\pi T)$ and $r=r(U,V)$. Note that the subscript `KS' represents the Kruskal-Szekeres. It is clear that the metric written in Kruskal coordinates is manifestly smooth at both the $U = 0$ and $V = 0$ horizons where $r(0,V)=r(U,0)=r_h$.

Now, we throw a particle into the black hole along the $V=0$ horizon at a boundary time $t$ in the past. The perturbation is negligible when the release time $t$ is short. Nevertheless, when $t$ is long enough, the particle's energy will exponentially boosted at $t=0$ slice and localised at the $V=0$, yielding
\begin{align}
\delta T^S_{VV} \sim \frac{1}{\sqrt{-g}} E_0 e^{\frac{2\pi}{\beta} t} \delta(V) \delta(\theta-\theta_1)\delta(\phi-\phi_1)\delta(\tilde\psi -\tilde\psi_1) \,,
\end{align}
where $E_0$ denotes the initial energy of the particle at the boundary, and $(\theta_1,\phi_1,\tilde\psi_1)$ represents the spatial coordinates at which the particle is released. Consequently, the initial perturbation becomes significant and leads to a strong backreaction on the geometry after the scrambling time $t_\ast \sim \frac{\beta}{2\pi} \ln{N^2}$, where $N^2\propto G_N^{-1}$ captures the number of degrees of freedom in the theory. The resulting geometry can be generally expressed by a shift along $U$ direction $U\to U+ H(\theta,\phi,\tilde\psi)\Theta(V)$, where $H(\theta,\phi,\tilde\psi)$ denotes the profile of the gravitational shock wave and $\Theta(V)$ is the Heaviside step function~\cite{Shenker:2013pqa,Dray:1984ha,Sfetsos:1994xa}. Redefining the $U,V$ coordinates, the back-reacted geometry becomes
\begin{align}
\begin{split} \label{eq:shockwaveG}
ds^2_{\text{after}} & =ds^2_{KS} +\frac{F}{4\pi^2T^2 UV} H(\theta,\phi,\tilde\psi) \delta(V) dV^2 -\frac{F_3 \left(\Omega_H -\Omega \right) }{ 8\pi T U} H(\theta,\phi,\tilde\psi) \delta(V) dV \tilde\sigma_3\,,\\
A_{\text{after}} & =A_{KS} -\frac{2A_t+\Omega_H A_\psi}{8 \pi T U} H(\theta,\phi,\tilde\psi) \delta(V) dV\,.
\end{split}
\end{align}
Plugging~\eqref{eq:shockwaveG} into the Einstein equation $\mathcal{E}_{ab}=8\pi G_N \delta T^S_{VV}$ and using $V\delta'(V)=-\delta(V)$ and $V\delta(V)=0$ at the horizon, one finds the equation for $H=H(\theta,\phi,\tilde{\psi})$
\begin{align}\label{eq:shockwaveH}
 \left( \square +\lambda^{sw}_1 +\lambda^{sw}_2 \partial_{\tilde{\psi}} +\lambda^{sw}_3 \partial^2_{\tilde{\psi}} \right) H \propto  \frac{1}{\sin\theta} E_0e^{\frac{2\pi}{\beta} t} \delta(\theta-\theta_1) \delta(\phi-\phi_1) \delta(\tilde\psi-\tilde\psi_1)  \,,
\end{align}
where the $\square$ denotes one quarter of the Laplacian on the unit 3-sphere expressed via the Hopf fibration as
\begin{equation} \label{eq:square}
\square = \partial_\theta^2+ \cot\theta \partial_\theta + \csc^2\theta \partial_\phi^2 -2 \cot\theta \csc\theta \partial_{\tilde{\psi}} \partial_\phi + \csc^2\theta \partial_{\tilde{\psi}}^2  \,.
\end{equation}
The constants $\lambda^{sw}_1, \lambda^{sw}_2, \lambda^{sw}_3$ in~\eqref{eq:shockwaveH} are all evaluated at the black hole horizon
\begin{eqnarray}
\begin{split} \label{eq:lambdaSW}
&\lambda_1^{sw}= \frac{F_2}{4} \left( F_1''-\frac{8}{L^2} +\frac{F_1'}{4} \left( \frac{8F_2'}{F_2} -\frac{5F_3'}{F_3 } \right) \right)-\frac{A_\psi^2}{3F_2}  -\frac{F_3}{24} \left( 2A_t'+\Omega A_\psi' \right)^2 -\frac{F_3^2 \Omega'^2}{16} + \frac{F_3 \Omega'}{32} \mathcal{M} \,, \\
& \lambda_2^{sw} = -\frac{1}{4} \sqrt{F_2F_3} \Omega' -\frac{1}{4} \sqrt{\frac{F_2}{F_3} } \mathcal{M}  \,, \qquad \lambda_3^{sw} = \frac{F_2-F_3}{F_3} \,,
\end{split}
\end{eqnarray}
where 
\begin{equation} \label{eq:mmm}
\mathcal{M} =\left( A_t +\frac{\Omega}{2}A_\psi \right) \left[ \frac{F_3^{3/2} \Omega'}{8\pi T \sqrt{F_2} } \left( A_t' +\frac{\Omega}{2} A_\psi' \right) -A_\psi' \right]\Bigg{|}_{r=r_h} \,.
\end{equation}
Note that, $\mathcal{M}$ is simplified by employing~\eqref{eq:tem} in combination with the condition $F_1/F_0=F_3/F_2$ at the horizon. One finds that the parameters $\lambda_i^{sw}$ explicitly depend on the $U(1)$ gauge potentials, varying under the $U(1)$ transformation or with the choice of boundary condition for the gauge field. We will discuss this point with more details in Section~\ref{sec:5}.

To investigate the butterfly effect, we need to solve the shock wave equation~\eqref{eq:shockwaveH}. Our analysis herein parallels that for Myers-Perry-AdS$_5$ black holes~\cite{Amano:2022mlu} and see Appendix~\ref{app:shockwave} for more details. The exact shock wave profile $H(\theta,\phi,\tilde\psi)$ can be obtained using the Wigner D-functions $D^{\mathcal{J}}_{\mathcal{KM}}(\theta, \tilde\psi, \phi)$–the eigenfunctions of the Laplacian on 3-sphere $S^3$. These functions can be expressed as
\begin{equation}
D^{\mathcal{J}}_{\mathcal{KM}}(\theta, \tilde\psi, \phi) = d^{\mathcal{J}}_{\mathcal{KM}}(\theta) e^{i\mathcal{K}\tilde\psi+i\mathcal{M}\phi}\,,
\end{equation}
with total angular momentum $\mathcal{J}=(2n+1)/2, n\in\mathbb{Z}$ and quantum numbers $\mathcal{K},\mathcal{M} = -\mathcal{J}, -\mathcal{J}+1, \cdots, \mathcal{J}$. They satisfy the following eigenvalue equation
\begin{align}  \label{eq:wignerD}
    \Box D^{\mathcal{J}}_{\mathcal{KM}} + \mathcal{J}(\mathcal{J}+1) D^{\mathcal{J}}_{\mathcal{KM}} =0\,.
\end{align}
Using the eigenvalue equation~\eqref{eq:wignerD} and completeness relation for Wigner D-functions, the exact solution of the shock wave equation~\eqref{eq:shockwaveH} is given by
\begin{eqnarray} \label{eq:Hsol}
\begin{split}
H(\theta, \theta_1, \tilde\psi, \phi) &=E_0 e^{2\pi T t} \sum_{\mathcal{J}=0,1/2,1,\cdots}^{\infty} \,
\sum_{\mathcal{K}=-\mathcal{J}}^{\mathcal{J}} \,
\sum_{\mathcal{M}=-\mathcal{J}}^{\mathcal{J}} \frac{2\mathcal{J}+1}{16\pi^2}  \frac{d^{\mathcal{J}}_{\mathcal{KM}}(\theta) d^{\mathcal{J}*}_{\mathcal{KM}}(\theta_1) \, e^{i\mathcal{K}\tilde\psi+i\mathcal{M}\phi} } {\mathcal{J}(\mathcal{J}+1)-\lambda_1^{sw}-i \lambda_2^{sw} \mathcal{K} + \lambda_3^{sw} \mathcal{K}^2} \,,
\end{split}
\end{eqnarray}  
where the delta function is sourced at $(\theta_1,\tilde\psi_1=0, \phi_1=0)$.

For general parameters $(\theta,\theta_1,\tilde\psi, \phi)$, it is challenging to handle the infinite summation in~\eqref{eq:Hsol} and derive a simple analytical expression for the shock wave. However, similar to the calculation in the planar black hole case~\cite{Zhao:2025gej}, when the shock wave profile depends solely on the $\tilde\psi$-coordinate --corresponding to OTOC operators constrained to a Hopf circle with identical $(\theta,\phi)$ coordinates while separated in $\tilde\psi$ -- the computation of~\eqref{eq:Hsol} simplifies dramatically. Without loss of generality, we may employ the isometry to fix $\phi = \theta = 0$. For such configurations, in the large black hole limit ($r_h\gg L$), the shock wave profile becomes highly localized near the equatorial plane with $\theta\approx\theta_1 \approx 0$, yielding a simple expression for $H$, \emph{i.e.}
\begin{align}
\begin{split}
H (\alpha) = E_0 & e^{2\pi T t}  \int_0^{\infty} d\mathcal{J} \int_{-\mathcal{J}}^\mathcal{J} d\mathcal{K}  \frac{\mathcal{J}}{4\pi^2}  \frac{e^{i\mathcal{K}\alpha}}{\mathcal{J}^2-\lambda^{sw}_1 -i \lambda^{sw}_2 \mathcal{K} + \lambda^{sw}_3 \mathcal{K}^2}  \,,
\end{split}
\end{align}
where $\alpha\equiv \tilde\psi+\phi$. Since $\phi$ is fixed to $0$, we have $\alpha = \tilde{\psi}$. This integral can be evaluated via contour integration, which gives the following shock wave profile (see Appendix \ref{app:shockwave}):
\begin{align} \label{eq:Happ}
 H(\tilde\psi)=  \frac{  E_0 e^{ 2\pi T t}  }{4\pi \tilde\psi} \bigg[  e^{\tilde\psi k_-^{sw}} \Theta(-\tilde\psi) - e^{-\tilde\psi k_+^{sw}} \Theta(\tilde\psi) \bigg] \,,
\end{align}
with
\begin{equation}
 k_\pm^{sw} = \frac{\lambda_2^{sw} \mp \sqrt{ \left( \lambda_2^{sw} \right)^2 - 4\lambda_1^{sw} \left( 1+ \lambda_3^{sw} \right)}}{2 \left( \lambda_3^{sw}+1 \right)}\,.  
\end{equation}
In the co-rotating coordinates, the angular coordinate is defined as $\tilde\psi=\psi-\Omega_H t$, where $t$ and $\psi$ represent the time and angular coordinates of the boundary field theory, respectively. In terms of $(t,\psi)$, the shock wave profile takes
\begin{align} \label{eq:swprofile}
 H(t,\psi) & =
\begin{cases}
-\frac{1}{4\pi |\psi-\Omega_H t| } e^{ 2\pi T_{+}   \left( t - \frac{L\psi}{2 v_{B,sw}^{+} } \right) } & \text{if } \psi-\Omega_H t > 0\,, \\
- \frac{1}{4\pi |\psi-\Omega_H t|} e^{ 2\pi T_{-}   \left( t - \frac{L\psi}{2 v_{B,sw}^{-} } \right) } & \text{if } \psi-\Omega_H t < 0\,,
\end{cases}
\end{align}
with $2\pi T_{\pm}=2\pi T\pm \Omega_H k_{\pm}^{sw}$. Here, we have taken into account that the spatial distance along the $\psi$-direction is taken to be $L\psi/2$, as indicated by~\eqref{eq:sbdy}. Therefore, the butterfly velocities can be read as
\begin{equation} \label{eq:vb}
v_{B,sw}^\pm = \frac{2\pi T_{\pm} L }{ 2 k_\pm^{sw} }  \,,
\end{equation}
where $\pm$ represent the information spreading parallel and antiparallel to the $\psi$ coordinate, respectively. Before continue, we parameterize the OTOC of~\eqref{eq:otoc} as $f(t,\vec{x})=1 -\epsilon H(t,\vec{x})$. In the present case one has $f(t,\psi)=1 -\epsilon H(t,\psi)$ with $H(t,\psi)$ given by~\eqref{eq:swprofile}. The Lyapunov exponent associated with the OTOC can be extracted from $f(t,\psi)$ at $\psi=0$ and $t>0$. One then finds that the Lyapunov exponent is given by $\lambda_L=2\pi T_+=2\pi T +\Omega_Hk_+^{sw}$ when $\Omega_H<0$ and by $\lambda_L=2\pi T_-= 2\pi T- \Omega_H k_-^{sw}$ when $\Omega_H>0$. Moreover, one can verify that $f(t,\psi)$ saturates the following bound
\begin{equation}
\frac{|\left(\partial_t +\Omega_H \partial_\psi\right)f(t,\psi)|}{1-f(t,\psi)} \leq 2\pi T \,,
\end{equation}
in the rotating ensemble generated by the positive operator $\theta_aQ_a=\beta\left(H+\Omega_H J\right)$, where $\theta_a$ denotes the chemical potential and $Q_a$ the corresponding generator~\cite{Mezei:2019dfv}.

\section{Pole-skipping phenomenon}\label{sec:4}

Studying the pole-skipping phenomenon requires analyzing near-horizon perturbations with ingoing boundary conditions, which can be effectively implemented using the infalling Eddington-Finkelstein coordinates $dv=dt+dr_\ast$. Furthermore, it is convenient to employ the co-rotating coordinate defined by $\tilde\psi=\psi-\Omega_Ht$. Therefore, in the Eddington-Finkelstein coordinates $(v,r,\tilde{\psi},\theta,\phi)$, the background fields take the form of
\begin{eqnarray}\label{eq:EFc}
\begin{split}
ds^2 &= -F dv^2 - \frac{F_3}{2} \left( \Omega - \Omega_H \right) \left(dv - \frac{dr}{ \sqrt{FF_1 }} \right) \tilde\sigma_3  +2\sqrt{ \frac{F}{F_1} } dv dr +\frac{F_2}{4} \left( \sigma_1^2+ \sigma_2^2 \right) + \frac{F_3}{4} \tilde\sigma_3^2\,,\\
A&= \left( A_t +\frac{\Omega_H}{2} A_\psi \right) dv + \frac{A_\psi}{2} \tilde\sigma_3 \,,
\end{split}
\end{eqnarray}
where we have used the gauge symmetry to set $A_r=-\frac{ 2A_t+\Omega_H A_\psi}{2\sqrt{FF_1}}$ in~\eqref{eq:EFc} to zero.\footnote{We have examined the calculation without using the radial gauge and found that the results of pole-skipping remain unchanged.}

Working in the ingoing Eddington-Finkelstein coordinates $(v,r,\tilde{\psi},\theta,\phi)$, the linearized perturbations propagating along the $v$ and $\tilde{\psi}$ direction can be expressed as
\begin{align} \label{eq:fluct}
h_{ab} =e^{-i\omega v +ik\tilde{\psi}} h_{ab} (r,\theta, \phi) \,, \qquad
a_b = e^{-i\omega v +ik\tilde{\psi}} a_{b} (r,\theta, \phi) \,,
\end{align}
where $a, b$ run in $\{v,r,\tilde{\psi},\theta,\phi\}$. Substituting the above perturbations into the linearized equations of motion, we will obtain a set of coupled partial differential equations. Similar to the Myers–Perry-AdS$_5$ black hole, the perturbations described above could be classified into tensor, vector, and scalar sectors using Wigner‑D functions, which are the eigenfunctions associated with the enhanced $U(2)\simeq SU(2)\times U(1)$ symmetry~\cite{Murata:2007gv,Murata:2008xr}.

Since the present work focuses on the pole-skipping phenomenon, it is sufficient to analyze the near-horizon properties of the fluctuation equations. A detailed analysis of the linearized equations shows that the $vv$ component of the Einstein equation takes the following form.
\begin{equation} \label{eq:ps00}
\begin{split}
& \bigg\{ \Big(\left( \partial_{\theta}^2 + \cot\theta \partial_{\theta} + \csc^2\theta \partial_{\phi}^2 - 2 i k \cot\theta \csc\theta \partial_{\phi} - \csc^2\theta k^2\right) +\lambda_1^{ps}+ i \lambda_2^{ps} k - \lambda_3^{ps} k^2  \Big) h_{vv}     \nonumber\\
& \quad -\frac{1}{F_3} \left( \omega - \frac{i F_1'}{2}\sqrt{\frac{F_2}{F_3}} \right) \bigg[ F_2 \big(\omega h_{\tilde\psi \tilde\psi}+ 2k h_{v \tilde\psi} \big)  -\cot^2\theta F_3 \Big( 2i \tan\theta h_{v\theta} +2k\sec\theta h_{v\phi} \nonumber\\
& \quad  -2k h_{v\tilde\psi} -\omega\tan^2\theta h_{\theta\theta} -\omega\sec^2\theta h_{\phi\phi} +2\omega \sec\theta h_{\phi\tilde\psi} -\omega h_{\tilde\psi \tilde\psi} \Big)    \nonumber\\
& \quad  -2i F_3 \Big( \partial_\theta h_{v\theta}+\csc^2\theta \partial_\phi h_{v\phi} -\cot\theta\csc\theta \partial_\phi h_{v\tilde\psi} \Big)  \bigg] \bigg\} \bigg|_{r=r_h} =0  \,.
\end{split}
\end{equation}
It is worth noting that the horizon equation presented above does not contain perturbations of the gauge field. It is obvious that the energy density perturbation $h_{vv}(r,\theta,\phi)$ decouples from other perturbations at the black hole horizon when
\begin{equation}
\omega =\frac{i F_1'}{2}\sqrt{\frac{F_2}{F_3}} = \frac{i F_0'}{2}\sqrt{\frac{F_1}{F_0}} = i 2\pi T\,,    
\end{equation}
corresponding to the well-known pole-skipping condition~\cite{Blake:2018leo}.
In particular, the $vv$ component of linearized Einstein equations at the horizon takes the form of
\begin{eqnarray} \label{eq:ps}
\begin{split}
\Big[ \partial_{\theta}^2 + \cot\theta \partial_{\theta} + \csc^2\theta \partial_{\phi}^2 - 2 i k \cot\theta \csc\theta \partial_{\phi}  - \csc^2\theta k^2 +\lambda_1^{ps}+ i \lambda_2^{ps} k - \lambda_3^{ps} k^2  \Big] h_{vv} = 0 \,,
\end{split}
\end{eqnarray}
with
\begin{eqnarray}
\begin{split} \label{eq:lambdaPS}
\lambda_1^{ps}= \lambda_1^{sw} -\frac{F_3\Omega'}{32}\mathcal{M} \,,\qquad  \lambda_2^{ps} = \lambda_2^{sw} +\frac{1}{4} \sqrt{ \frac{F_2}{F_3} } \mathcal{M}  \,, \qquad \lambda_3^{ps} = \lambda_{3}^{sw}\,,
\end{split}
\end{eqnarray}
evaluated at the horizon. At first sight, the pole-skipping equation~\eqref{eq:ps} does not coincide with the source-less shock wave equation~\eqref{eq:shockwaveH}, due to the difference between the parameters $\lambda_i^{sw}$ and $\lambda_i^{ps}$. However, if the $U(1)$ gauge field is turned off, these equations reduce to those of the Myers–Perry-AdS$_5$ black hole, and the equivalence between pole-skipping and the shock wave is recovered~\cite{Amano:2022mlu}.

To identify the pole-skipping point, we make use of the underlying $SU(2)\times U(1)$ symmetry and analyze~\eqref{eq:ps} employing the Wigner D-functions $D^{\mathcal{J}}_{\mathcal{KM}}(\theta, \tilde\psi, \phi)$ which satisfy the eigenvalue equation~\eqref{eq:wignerD}. Generally, for a specified $\mathcal{J},\mathcal{K},\mathcal{M}$, there exists a unique regular solution for the eigenvalue equation~\eqref{eq:wignerD}. Nevertheless, to make contact to the pole-skipping phenomenon, we will extend the quantum numbers $\mathcal{J},\mathcal{K},\mathcal{M}$ from rational numbers to complex numbers. Therefore, by comparing~\eqref{eq:ps} with~\eqref{eq:wignerD}, we know that at the pole-skipping point the quantized numbers $\mathcal{K}$ and $\mathcal{J}$ should satisfies
\begin{align} \label{eq:pskj}
\lambda_1^{ps}+i \lambda_2^{ps} \mathcal{K} - \lambda_3^{ps} \mathcal{K}^2 = \mathcal{J}(\mathcal{J}+1)\,,
\end{align}
where we have set $k=\mathcal{K}$. For an arbitrary choice of $\mathcal{J}$, the value of $\mathcal{K}$ can be determined from~\eqref{eq:pskj}. However, for simplicity, we consider case where $\mathcal{K}$ is parallel to the rotation direction and saturates its bounds, \emph{i.e.}, $\mathcal{K}=\mathcal{J}$.\footnote{The situation is a bit more elaborate when $\mathcal{K}\neq\mathcal{J}$, where the perturbations rotate transverse to the rotation of the background charged fluid. Investigating the pole-skipping and its relation to chaos under such conditions would be interesting. However, this lies beyond the scope of the present work and will be addressed in future work.}

Compared to the shock wave solution~\eqref{eq:Happ}, the equation~\eqref{eq:pskj} can be solved analytically without taking the large black hole limit. In this case, the corresponding value of $\mathcal{K}$ determined from~\eqref{eq:pskj} becomes a complex number with both non-vanishing real and imaginary parts, \emph{i.e.}
\begin{align} \label{eq:pskk}
 \mathcal{K} =\pm \frac{ \pm \left( i \lambda_2^{ps} -1 \right)+ i \sqrt{ \left( \lambda_2^{ps} +i \right)^2 - 4\lambda_1^{ps} \left(1+ \lambda_3^{ps} \right)}} { 2\left( 1+\lambda_3^{ps} \right)}  \,.
\end{align}
Such type of perturbations may be interpreted as a finite-size effects: when the spatial directions are compactified, perturbations propagating in opposite directions interfere and form spatially decaying and oscillatory waves, analogous to underdamped oscillations in mechanical system.

On the other hand, to compare with the shock wave calculation, it is necessary to consider the large black hole limit where $\mathcal{J} \sim \mathcal{K} \sim r_h/L \gg 1$. In this case, the right hand side of~\eqref{eq:pskj} can be approximated as $\mathcal{J}^2$, yielding
\begin{align} \label{eq:pskk}
 \mathcal{K} =\pm  i \frac{\lambda_2^{ps} \mp \sqrt{ \left( \lambda_2^{ps} \right)^2 - 4\lambda_1^{ps} \left(1+ \lambda_3^{ps} \right)}} { 2\left( 1+\lambda_3^{ps} \right)} = \pm i k_\pm^{ps}  \,,
\end{align}
which is purely imaginary. Therefore, it follows from~\eqref{eq:ps00} and~\eqref{eq:pskk} that the energy-density perturbation exhibits pole-skipping at $(\omega =i2\pi T,\, k= \pm ik_\pm^{ps})$. We emphasize that the above near-horizon analysis was performed in co-moving coordinates $(t,\tilde\psi,\theta,\phi)$, which are related to the standard boundary coordinates $(t,\psi,\theta,\phi)$ via the transformation $\tilde\psi =\psi -\Omega_H t$.
Under such transformation, the frequency gets modified while the momentum remains unchanged~\cite{Liu:2020yaf,Amano:2022mlu}, \emph{i.e.} $\omega_{bc} = \omega_{cm}+\Omega_H k_{cm} \,, k_{bc}=k_{cm}$. Therefore, the location of pole-skipping point written in the boundary coordinates is
\begin{equation} \label{eq:psbdy}
(\omega,\, k)=\left( \omega_\pm^{ps},\, k_\pm^{ps} \right) = \left( i\left(2\pi T \pm \Omega_H  k_\pm^{ps} \right),\, i k_\pm^{ps} \right)  \,.
\end{equation}
Consequently, the butterfly velocity extracted from the pole-skipping point is given by
\begin{equation} \label{eq:velocityPS}
v_{B,ps}^\pm = \frac{\omega_\pm^{ps} L }{ 2 k_\pm^{ps} } = \frac{ \left(2\pi T \pm \Omega_H  k_\pm^{ps} \right) \left( 1+\lambda_3^{ps} \right) L }{ \lambda_2^{ps} \mp  \sqrt{ \left( \lambda_2^{ps} \right)^2 - 4\lambda_1^{ps} \left(1+ \lambda_3^{ps} \right)}} \,.
\end{equation}
In the following, we discuss $v_B^\pm$ in several limiting cases.
\begin{itemize}
\item Static limit: This corresponds to setting $\Omega=A_\psi=0$. The solution to field equations~\eqref{eq:eom} is the Reissner-N\"ordstorm-AdS (RN-AdS) black hole, with
\begin{equation}
F_0=F_1= 1+ \frac{r^2}{L^2} -\frac{2(m-q)}{r^2} +\frac{q^2}{r^4} \,,\quad  A_t=  -\frac{\sqrt{3} q }{ r^2 } \,,\quad  F_2=F_3=r^2 \,.
\end{equation}
In the large black hole limit $r_h/L\gg 1$, the butterfly velocity reads
\begin{equation} \label{eq:RNvB}
v_B^{RN} = v_B^{Sch} \sqrt{ 1 -\frac{q^2L^2}{2r_h^6} +\frac{L^2}{2r_h^2} } \,,
\end{equation}
where $v_B^{Sch}= \pm \sqrt{2/3}$ is the butterfly velocity for the planar Schwarzschild black hole.\footnote{As previously discussed, when the large black hole limit is not taken, the butterfly velocity $v_B$ computed from~\eqref{eq:pskj} is given by
\begin{equation}\label{vbRN}
\tilde{v}_B^{RN} = v_B^{Sch} \sqrt{ 1 -\frac{q^2L^2}{2r_h^6} +\frac{L^2}{3r_h^2} }  -\frac{i L}{3r_h}\,.
\end{equation}
When $r_h$ is comparable to the AdS radius $L$, the butterfly velocity $\tilde{v}_B^{RN}$ acquires an imaginary component. The interpretation of a complex butterfly velocity is subtle, although it may be understood as a manifestation of finite-size effects for black holes with spherical topology. However, as $r_h$ becomes significantly larger than $L$, the value of $\tilde{v}_B^{RN}$ approaches $v_B^{RN}$ that obtained in the large black hole limit.}

\item Neutral limit: In this case, the background reduces to the Myers–Perry-AdS$_5$ black hole~\cite{Hawking:1998kw,Amano:2022mlu}. The corresponding butterfly velocity is investigated in~\cite{Amano:2022mlu}.
\end{itemize}
The butterfly velocity in~\eqref{eq:velocityPS} incorporates both the rotation and electric charge, demonstrating the significant influence of the gauge field on information propagation through the system. For the pole-skipping case, a similar expression for the Lyapunov exponent can be obtained from~\eqref{eq:psbdy} and~\eqref{eq:greenps} but with $k_\pm^{sw}$ replaced by $k_\pm^{ps}$, \emph{i.e.} $\lambda_L=2\pi T \pm\Omega_Hk_\pm^{ps}$. Recalling that our primary goal in this work is to explore the connection between pole-skipping and quantum chaos in rotating fluids with non-trivial matter distributions. 
Therefore, we will address the discrepancy between the standard OTOC and the pole-skipping results in the next section.

\section{On the equivalence between OTOC and pole-skipping}\label{sec:5}

Our analysis reveals an apparent inconsistency between the two approaches to quantum chaos, see~\emph{e.g.} the difference of butterfly velocities in~\eqref{eq:vb} and~\eqref{eq:velocityPS}. The key point is the contribution due to the difference between the parameters $\lambda_i^{sw}$ and $\lambda_i^{ps}$ in the presence of $U(1)$ gauge field. We will show that this unexpected issue can be solved by considering an appropriate boundary condition for the gauge potential at the event horizon.

We begin with the gauge dependence of the shock wave equation, discussing how the $U(1)$ gauge of the system can be appropriately fixed. First, for rotating black holes without electric charge or charged black holes without rotation, it can be readily shown that $\mathcal{M}=0$, see~\eqref{eq:mmm}. In such cases, the shock wave equation becomes invariant under the $U(1)$ gauge transformation, and the issue of gauge dependence is automatically eliminated. The source-free part of the shock wave equation~\eqref{eq:shockwaveH} exactly matches the pole-skipping equation~\eqref{eq:ps}, demonstrating that both pole-skipping and OTOC provide equivalent descriptions of quantum chaos in holographic CFTs~\cite{Amano:2022mlu}.

In contrast, for the charged and rotating black hole studied in this work, the issue of $U(1)$ gauge dependence becomes unavoidable, for example, $\lambda_i^{sw}$ of~\eqref{eq:lambdaSW} explicitly depend on $A_t$ and $A_\psi$, varying under the $U(1)$ transformation. On the other hand, the butterfly velocity obtained from the shock wave equation is a physical observable and should therefore be gauge-independent. This implies that, an appropriate boundary condition must be imposed to fix the $U(1)$ gauge freedom of the system when solving the shock wave equation~\eqref{eq:shockwaveH}.

We find that the Wilson loop argument yields a clearly physical boundary condition for the gauge potential at the black hole event horizon, providing a natural way to fix the $U(1)$ gauge freedom of the system. This idea was introduced in~\cite{Horowitz:2010gk} for holographic superconductors, \emph{i.e.} a static black hole with charged scalar hair. We now generalize this argument to the rotating case, which ensures $\mathcal{M}=0$, thereby guaranteeing the consistency between the OTOC and pole-skipping results. 

We begin with the black hole background in the co-rotating coordinates~\eqref{trmetric}, which bears the closest resemblance to the static black hole near the event horizon of~\cite{Horowitz:2010gk}. For the description of thermal properties, we should consider the Euclidean solution which is obtained via analytic continuation of the Lorentzian metric~\eqref{trmetric} by substituting $t \to -i\tau_E$ and $\Omega_H \rightarrow i\Omega_{H,\text{E}}$. Regularity at the horizon $r_h$ requires the identifications
\begin{equation}
\tau_\text{E} \sim \tau_\text{E} +\beta \,,
\end{equation}
where $\beta=1/T$ is the inverse temperature. In the resulting coordinates $(\tau_\text{E},\tilde\psi_\text{E})$, the temporal component of the gauge field becomes $A_{\tau_\text{E} } |_{Euclidean} = -i \left( A_t+\frac{\Omega_H}{2}A_\psi\right)|_{Lorentzian}$. Since the Wilson loop of $A_\mu$ around the Euclidean time circle must remain finite and gauge invariant~\cite{Horowitz:2010gk}, the condition $A_{\tau_\text{E} }=0$ is enforced at the horizon. Otherwise, the Wilson loop is nonzero around a vanishing circle, resulting in that the Maxwell field is singular. As a consequence, the term $\mathcal{M}$ of~\eqref{eq:mmm} vanishes identically, thereby establishing exact equivalence between pole-skipping phenomenon and OTOC. Therefore, we can safely conclude that the two butterfly velocities are equal,~\emph{i.e.}
\begin{equation} \label{eq:butterfly}
v_{B,sw}^\pm =v_{B,ps}^\pm \equiv v_B^\pm \,.
\end{equation}
Likewise, after fixing the boundary condition of the gauge potential, the Lyapunov exponents computed via the two methods are found to coincide.

Additionally, setting the temporal component of the gauge field to zero in the co-rotating frame is consistent with the conventional gauge choice employed in static backgrounds. The boundary condition $ A_t(r_h)+\frac{\Omega_H}{2}A_\psi(r_h)=0$ through the Wilson loop plays an crucial role in evaluating the on-shell Euclidean action and free energy, ultimately yielding the expected form of the first law of thermodynamics. Only within this boundary condition can consistent black hole thermodynamics be obtained through holographic renormalization, thereby further highlighting the essential role of imposing a physical boundary condition for the $U(1)$ gauge potential at the event horizon (see Appendix~\ref{app:clp} for more details).

\section{Butterfly velocity in CLP black hole}\label{sec:6}
The analysis presented above applies to five-dimensional charged rotating black holes with equal angular momenta. However, in general, for either $k=0$ or arbitrary non-zero $k$, few analytical solutions are available in literature. Nevertheless, the system admit charged rotating black hole known in closed form: the CLP solution when $k=k_{susy}=2/\sqrt{3}$~\cite{Cvetic:2004hs,Madden:2004ym}. Thus, to explicitly demonstrate our results, we perform a direct computation of butterfly velocity in the CLP solution which is dual to strongly coupled charged rotating fluids.

The expression of the CLP black hole is given by~\cite{Madden:2004ym,Blazquez-Salcedo:2016rkj}
\begin{equation}\label{eq:CLP}
\begin{split}
F_1(r) &= 1-\frac{2m(1-\frac{a^2}{L^2})-2q}{r^2}+\frac{2a^2m+(1-\frac{a^2}{L^2})q^2}{r^4}+\frac{r^2}{L^2}, \\
F_2(r) &=r^2\,, \qquad F_3(r) = r^2({1+\frac{2a^2m}{r^4}}-\frac{a^2q^2}{r^6}) \,,\qquad 
F_0(r) =\frac{F_1(r)} {1+\frac{2a^2m}{r^4} -\frac{a^2q^2}{r^6}} \,,  \\
\Omega(r) &= \frac{ 2 a(2m-q-\frac{q^2}{r^2})}{r^4 (1+\frac{2a^2m}{r^4}-\frac{a^2q^2}{r^6})} \,, \qquad A_t(r) = -\frac{\sqrt{3}q}{r^2} +\xi \,,\qquad  A_\psi(r)= \frac{\sqrt{3}aq} {r^2} \,,
\end{split}
\end{equation}
where $m, a$ and $q$ are parameters related to the mass, rotation and electrical charge of the system. The AdS boundary is located at $r\to\infty$. See Appendix~\ref{app:clp} for more details on the CLP solution. Moreover, the constant $\xi$ in $A_t$ is introduced to ensure that $A_t+\frac{\Omega_H}{2} A_\psi$ vanishes at the event horizon $r_h$ as discussed in Section~\ref{sec:5}.\footnote{Alternative methods exist for fixing $\xi$, such as imposing the regularity condition by requiring the contraction of the Killing vector $\chi =\partial_t + \Omega_H\partial_{\psi}$ with the gauge field to vanish at the horizon: $\chi^b A_b \big|_{r=r_h}=0$~\cite{Cassani:2019mms,Cassani:2022lrk}. The value of $\xi$ determined in this approach is identical to ours.} This determines $\xi$ as
\begin{equation}\label{eq:CLPgauge}
\begin{split}
\xi = \frac{\sqrt{3} q\left(r_h^4 +a^2 q \right) }{ r_h^6 -a^2 \left( q^2 -2m r_h^2 \right) } \,.
\end{split}
\end{equation}
Crucially, only with this specific choice do the OTOC calculations become fully consistent with the pole-skipping analysis. This agreement reflects the consistency between Wilson loop and the framework of quantum chaos.

\begin{figure}[ht]
\begin{center}
\includegraphics[width=0.46\textwidth]{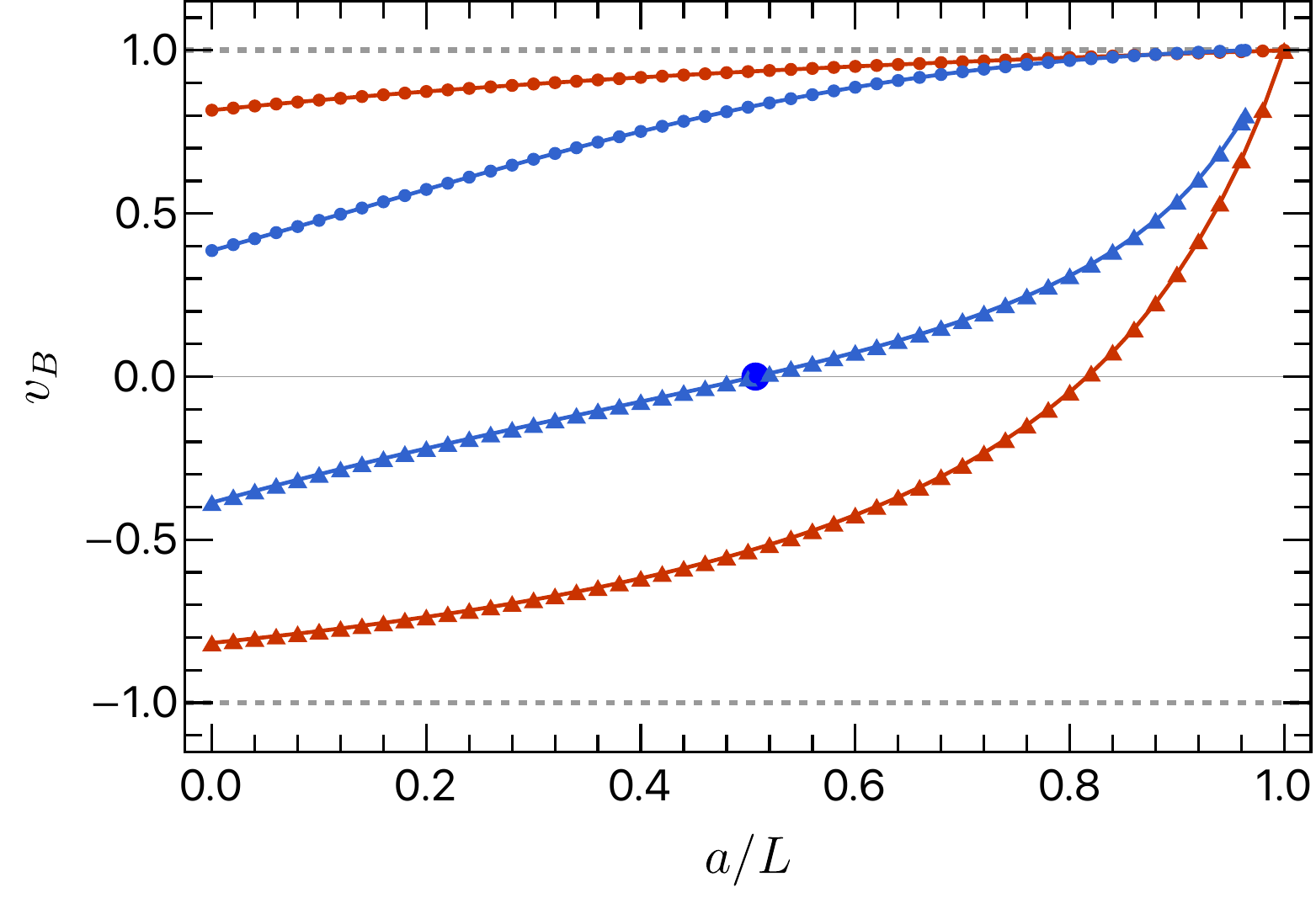}
\includegraphics[width=0.46\textwidth]{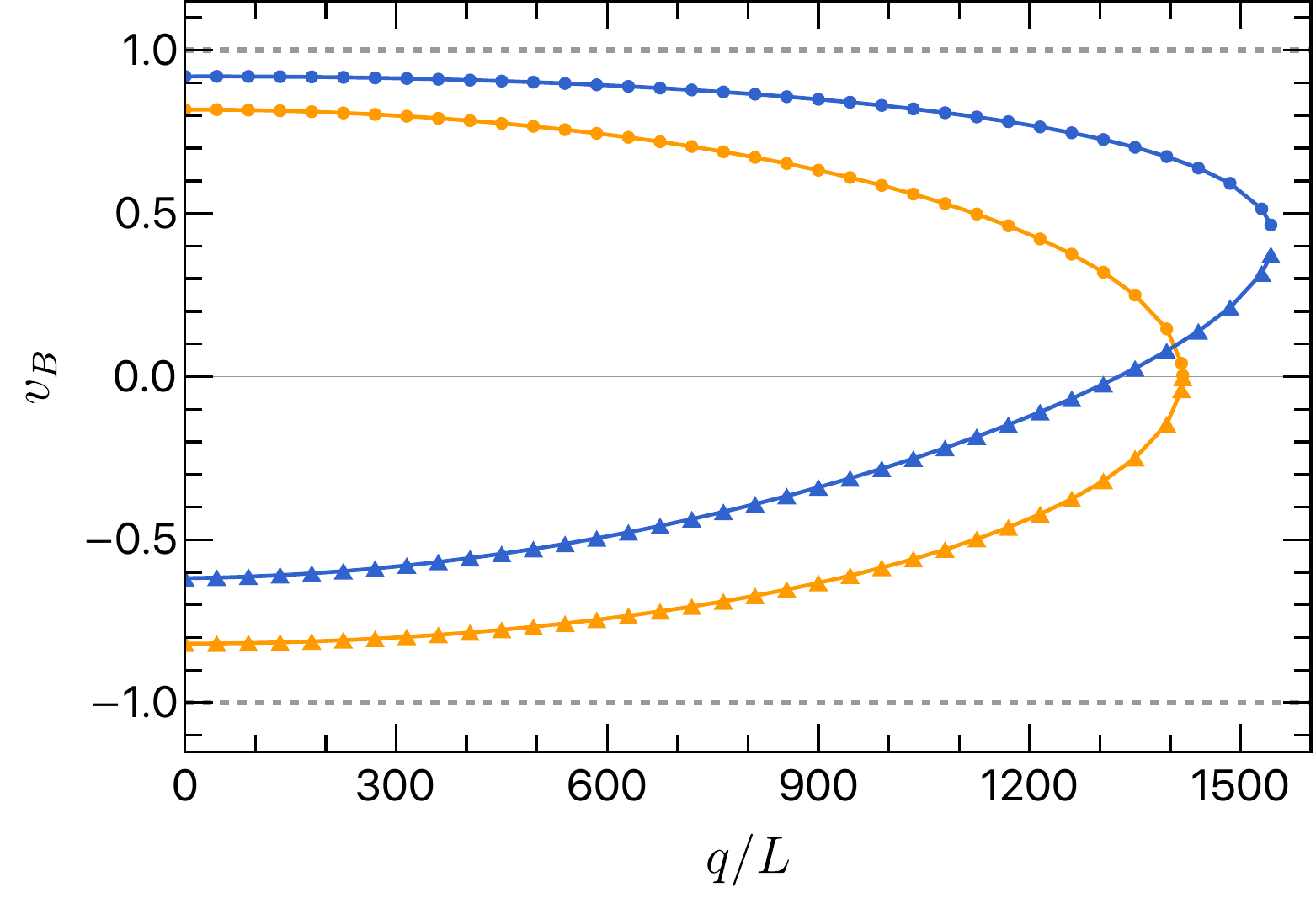}
\end{center}
\vspace{-0.3cm}
\caption{\small Left: Butterfly velocity $v_B^\pm$ of~\eqref{eq:butterfly} as a function of the rotation parameter $a$ for the CLP solution with $q=1250$ (blue curve) and for the Myers-Perry-AdS$_5$ (red curve). Right: Butterfly velocity $v_B^\pm$ as a function of the charge parameter $q$ for the CLP solution with $a=0.4$ (bule curve) and for the RN-AdS (yellow curve). In both polts, dotted markers correspond to the $``+"$ sector of $v_B$, while triangle markers correspond to the $``-"$ sector. The gray dashed lines indicate the speed of light. The large black hole limit is ensured by setting $r_h=10,\, L=1$.}\label{fig:vB}
\end{figure}

There are several interesting limiting cases for CLP black holes:
\begin{itemize}
\item
For $q=0$ (and thus $A_t=A_\psi=0$), the CLP black hole reduce to the rotating Myers-Perry-AdS$_5$ black holes with equal angular momentum $J_1=J_2=J$~\cite{Hawking:1998kw}. 
\item 
Setting $a=0$ (implying $\Omega =A_\psi=0$) yields the electrically charged Reissner-Nordstr\"om (RN) AdS$_5$ black hole. 
\end{itemize}
In both cases, $\mathcal{M}$ vanishes in~\eqref{eq:lambdaSW} and~\eqref{eq:lambdaPS} automatically, implying $\lambda_i^{s}=\lambda_i^{ps}$~\cite{Amano:2022mlu,Blake:2016jnn}.

\begin{figure}[ht]
\begin{center}
\includegraphics[width=0.6\textwidth]{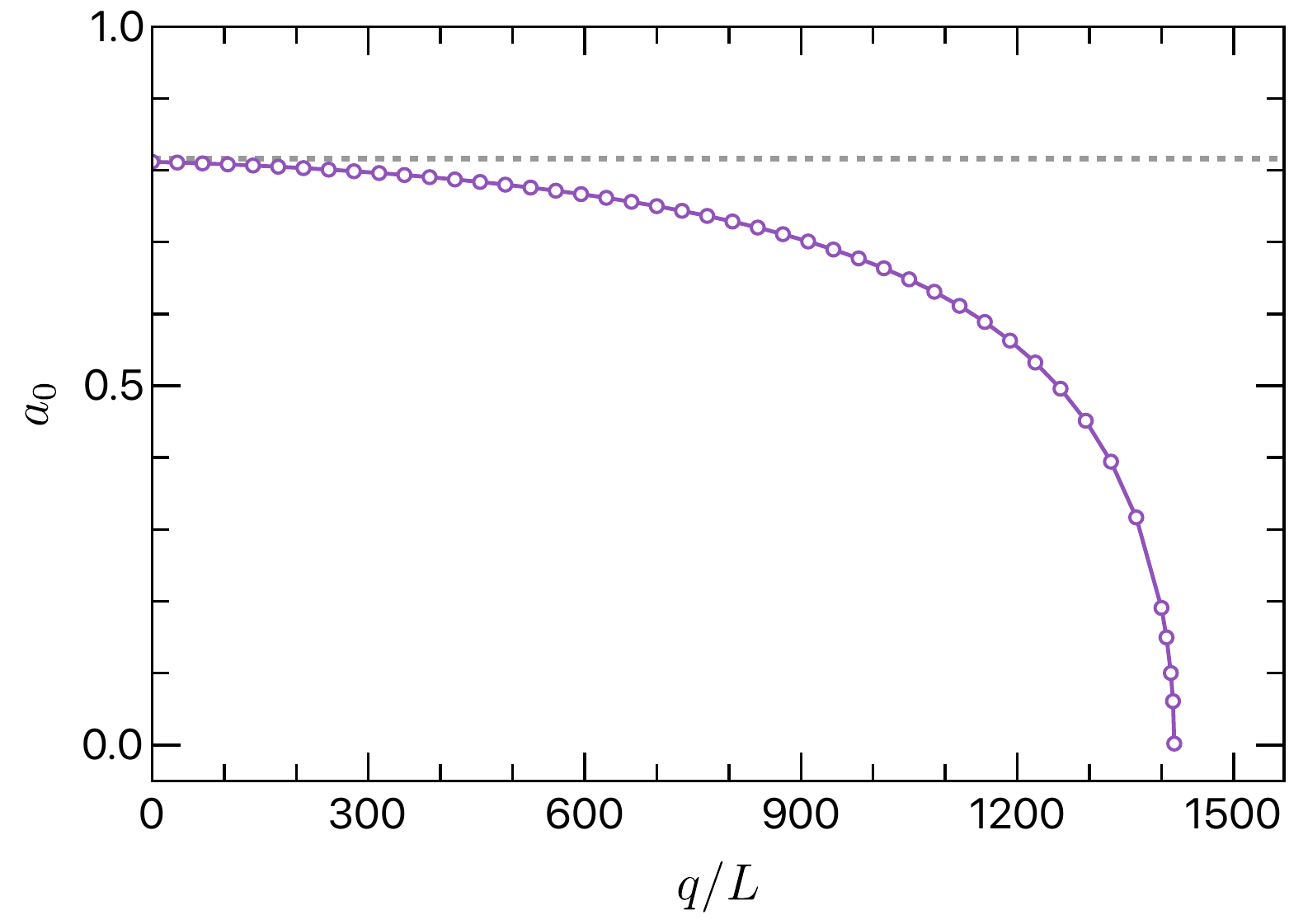}
\end{center}
\vspace{-0.3cm}
\caption{\small 
The purple curve shows the root $a_0$ of $v_B^-=0$ as a function of $q$ for the CLP black hole, while the dashed line corresponds to that of the Myers-Perry-AdS$_5$ black hole with $a_0=\sqrt{2/3}$. We have set $r_h=10, L=1$.}\label{fig:vB0}
\end{figure}

Since the analytical derivation of butterfly velocity in~\eqref{eq:vb} and~\eqref{eq:velocityPS} assumes the large black hole limit~\emph{i.e.} $r_h\gg L$, we explicitly present our results within this regime. In our numerical calculations, we take $r_h=10L$ and the corresponding results are shown in Figure~\ref{fig:vB}. Furthermore, physical constraints such as the non-negativity of temperature ($T \geq 0$) require $a/L < 1$ in the CLP black hole. It is worth noting that the condition $a/L < 1$ also appears in the Myers–Perry–AdS$_5$ black hole when higher-order terms in the expansion of butterfly velocity are included. As shown by the blue curves in the left panel of Figure~\ref{fig:vB}, in the absence of rotation, $v_B^+$ and $v_B^-$ are symmetric and are equal to the result~\eqref{eq:RNvB} for the RN-AdS black hole. As rotation is introduced, the butterfly velocity $v_B^+$ (blue dotted curve) increases monotonically from its static RN-AdS value. In contrast, the magnitude of $v_B^-$ (blue triangle curve) first decreases to zero with increasing the rotation parameter $a/L$, then $v_B^-$ becomes positive and continues to grow as the rotation parameter is increased further. However, unlike the case of the Myers-Perry-AdS$_5$ black hole, although $a/L$ cannot reach $1$, we observe that $v_B^+$ gradually approaches $1$, while $v_B^-$ can only increase to a value less than 1. Moreover, due to the presence of charge $q$, the zero of $v_B^-$ occurs at a smaller value of $a/L$ compared with the case of Myers-Perrry-AdS$_5$ black hole. For the parameters consider here, the location where $v_B^-$ vanishes is approximately $a/L\approx 0.507$ (marked by the blue dot in the left panel of Figure~\ref{fig:vB}).

Additionally, the right panel of Figure~\ref{fig:vB} compares the behavior of butterfly velocities in the CLP solution (with $a=0.4$) and RN-AdS black hole with respect to the charge parameter $q$. It is observed that the presence of rotation breaks the equality between the magnitudes of the butterfly velocities in the directions parallel and antiparallel to the $\psi$ direction. As the charge $q$ increases, $v_B^+$ (blue dotted curve) decreases monotonically similar to the behavior in the RN-AdS case, yet it remains positive at large $q$. On the other hand, the behavior of $v_B^-$ (blue triangle curve) resembles that observed when increasing the rotation parameter $a$ at a fixed $q$ (see the left panel of Figure~\ref{fig:vB}): starting from a negative value, $v_B^-$ crosses zero and becomes positive with the increasing of $q$. Denoting the root of $v_B^-(a)=0$ as $a_0$, we show the behavior of $a_0$ as a function of $q$ in Figure~\ref{fig:vB0}. One can see that as $q$ increases from zero to its maximum value, $a_0$ decreases monotonically to zero.

\section{Discussion and outlook}\label{sec:7}
We have investigated quantum chaos in holographic conformal field theories dual to charged rotating black holes in the Einstein–Maxwell–Chern–Simons theory~\eqref{eq:action}. The Lyapunov exponent $\lambda_L$ and butterfly velocity $v_B$ are computed using both pole-skipping analysis and the shock wave method. Interestingly, although the energy density perturbation exhibits the pole-skipping phenomenon precisely at $\omega=i2\pi T$, the resulting pole-skipping equation~\eqref{eq:ps} does not coincide with the source-less version of shock wave equation~\eqref{eq:shockwaveH}. Particularly, we found that the results of shock wave equation~\eqref{eq:shockwaveH} depend explicitly on the value of gauge potential at the event horizon. We have solved this issue by imposing a physical boundary condition for the gauge potential at the horizon, yielding perfect agreement between the OTOC and the pole-skipping results. This result further demonstrates the equivalence between pole-skipping and shock wave methods in rotating spacetimes with nontrivial matter fields, providing additional support to the idea that the hydrodynamic nature of quantum chaos may serve as a defining feature of maximally chaotic systems~\cite{Blake:2017ris,Blake:2021wqj,Knysh:2024asf}.

Furthermore, in the supergravity case $k=k_{susy}=2/\sqrt{3}$, the butterfly velocity is explicitly computed in the large black hole limit using the CLP solution, as shown in Figure~\ref{fig:vB}. As $a/L$increases, $v_B^+$ increases monotonically and approaches $1$ at a value of $a/L$ smaller than $1$, while $v_B^-$, starting from a negative value, crosses zero and subsequently increases to a value less than $1$ (see~\emph{e.g.} the left panel of Figure~\ref{fig:vB}). These properties of $v_B^\pm$ show some similarities with the Myers-Perry-AdS$_5$ black hole. However, the presence of charge significantly shifts the zero of $v_B^-$ to smaller values of $a/L$, see Figure~\ref{fig:vB0}. Since our analysis incorporates all relevant contributions in the large black hole limit, the results can not be simply explained from the perspective of a boosted Schwarzschild-AdS$_5$ black hole~\cite{Amano:2022mlu}. Finally, we have tried to derive a generalization of the identity $T_{vr} h_{vv} -\delta T_{vv}=0$ to the rotating spacetime. It was found that no simple expression exists. Thus, we conducted a direct examine of the identity $T_{vr} h_{vv} -\delta T_{vv}=0$ in the CLP solution, which revealed that the equation does not hold. This implies that even when this identity is not satisfied, the equivalence between pole-skipping and OTOC remains valid for charged rotating black holes.

There are several promising research directions valuable to explore. First, a direct numerical investigation of energy density perturbations in CLP black holes is necessary to verify that the complex-valued dispersion relation for sound modes does pass through the chaos point. Second, butterfly effects can be alternatively probed via entanglement wedge reconstruction~\cite{Mezei:2016wfz,Dong:2022ucb,Baishya:2024gih,Lilani:2025wnd}. A comparison between the present methods and the entanglement wedge analysis for quantum chaos would yield more insights into the maximal chaotic system~\cite{Chua:2025vig}. Third, studying the higher-order pole-skipping points and pole-skipping in other types of fields would be interesting~\cite{Blake:2019otz,Pan:2024azf}. Finally, the present work can be directly extended to charged rotating black holes in five-dimensional $U(1)^3$ gauged $\mathcal{N}=2$ supergravity coupled to two vector multiplets~\cite{Cvetic:2004ny,Madden:2004ym} or higher dimensional rotating and equal charge black holes~\cite{Chow:2007ts}. Further generalization to EMCS theories with general matter content is also possible \emph{i.e.} rotating charged hairy black holes~\cite{Dias:2024edd}. Moreover, the results presented herein should remain valid in the absence of Chern-Simons terms, despite the lack of analytical Kerr-Newman-AdS$_5$ solutions. Thus, it is possible that this phenomenon appears in Kerr-Newman-AdS$_4$ black holes.

It would be valuable to further identify the horizon symmetries in rotating black holes that correspond to the shift symmetry postulated in the effective field theory description of maximally chaotic systems~\cite{Blake:2017ris,Knysh:2024asf}, and to examine their relationship with both OTOC calculations and the pole-skipping phenomenon. It will be interesting to give a general proof of the equivalence between pole-skipping and OTOC in rotating black holes. A key observation is that the properties of quantum chaos primarily depend on the physics at the black hole horizon. This suggests that for generic rotating black holes, a coordinate transformation making the metric diagonal at the event horizon could potentially enable the application of arguments originally developed for static configurations~\cite{Blake:2018leo} to be used to demonstrate this equivalence. Nevertheless, the validity of this proposal remains to be rigorously established. In addition to the perspective of the Wilson loop, our findings highlight that imposing a physical boundary condition for the gauge potential at the event horizon is crucial, and it is imperative to explore alternative viewpoints to achieve a deeper understanding. Finally, away from the large black hole limit, the  butterfly velocity could take a complex value, see \emph{e.g.}~\eqref{vbRN}. This feature is due to the finite-size effect as it disappears in the large black hole limit and is a robust phenomenon, irrespective of the matter and rotation. This suggests that the OTOC~\eqref{eq:otoc} might develop an spatially modulation structure for a strongly coupled system on 3-sphere $S^3$.  It is desirable to understand the physical meaning of this feature.

\acknowledgments
We thank Liang Ma, Yan Liu, Ya-Wen Sun, Wen-Bin Pan and Jie-Qiang Wu for useful discussions. This work was supported by the National Natural Science Foundation of China Grants No.\,12525503, No.\,12505087, No.\,12247156, No.\,12447134, No.\,12447101 and Postdoctoral Innovation Project of Shandong Province SDCX-ZG-202503036.

\appendix
\section{Solution to shock wave equation}\label{app:shockwave}
We provide more details on the computation of the shock wave equation~\eqref{eq:shockwaveH}. Since the background spacetime possesses the same $SU(2)\times U(1)$ symmetry as the Myers-Perry-AdS$_5$ black hole with equal angular momentum, the method for solving the shock wave equation is therefore adapted from the Myers-Perry-AdS$_5$ case~\cite{Amano:2022mlu}. The exact solution of the shock wave equation can be obtained using the Wigner D-functions $D^{\mathcal{J}}_{\mathcal{KM}}(\theta, \tilde\psi, \phi)$--the eigen-functions of the Laplacian on $S^3$. 
The Wigner D-functions satisfy the following eigenvalue equations.
\begin{align}  \label{eq:app:wignerD}
    \Box D^{\mathcal{J}}_{\mathcal{KM}} + \mathcal{J}(\mathcal{J}+1) D^{\mathcal{J}}_{\mathcal{KM}} =0 \,,
\end{align}
where $\mathcal{J}=(2n+1)/2, n\in\mathbb{Z}$ the total angular momentum and the quantum number $|\mathcal{K}| \leq \mathcal{J}, |\mathcal{M}| \leq \mathcal{J}$. Similar to the case of spherical harmonics $Y^m_l$ for $S^2$, the Wigner D-functions form a complete set on $S^3$. The completeness relation for the Wigner D-functions reads
\begin{equation} \label{eq:app:completeR}
\frac{\delta(\theta-\theta_1) }{\sin\theta} \delta(\phi-\phi_1) \delta(\tilde\psi-\tilde\psi_1) = \sum_{\mathcal{J}=0,1/2,1,\cdots}^{\infty} \sum_{\mathcal{K}=-\mathcal{J}}^{\mathcal{J}} \sum_{\mathcal{M}=-\mathcal{J}}^{\mathcal{J}} \frac{2\mathcal{J}+1}{16\pi^2} D^{\mathcal{J}}_{\mathcal{KM}}(\theta, \tilde\psi, \phi) D^{\mathcal{J}*}_{\mathcal{KM}}(\theta_1, \tilde\psi_1, \phi_1) \,.\nonumber
\end{equation}
Using~\eqref{eq:app:wignerD} and the completeness relation, the analytical shock wave solution is expressed as
\begin{eqnarray} \label{eq:app:Hsol}
\begin{split}
H(\theta, \theta_1, \tilde\psi, & \phi) =E_0 e^{2\pi T t} \sum_{\mathcal{J}=0,1/2,1,\cdots}^{\infty} \,
\sum_{\mathcal{K}=-\mathcal{J}}^{\mathcal{J}} \,
\sum_{\mathcal{M}=-\mathcal{J}}^{\mathcal{J}} \frac{2\mathcal{J}+1}{16\pi^2}  \frac{d^{\mathcal{J}}_{\mathcal{KM}}(\theta) d^{\mathcal{J}*}_{\mathcal{KM}}(\theta_1) \, e^{i\mathcal{K}\tilde\psi+i\mathcal{M}\phi} } {\mathcal{J}(\mathcal{J}+1)-\lambda_1^{sw}-i \lambda_2^{sw} \mathcal{K} + \lambda_3^{sw} \mathcal{K}^2} \,.
\end{split}
\end{eqnarray}
where the normalized delta function is sourced at $(\theta_1, \tilde\psi_1=0, \phi_1=0)$.

For general parameters $(\theta,\theta_1,\tilde\psi, \phi)$, handling the infinite summation in~\eqref{eq:app:Hsol} to derive a simple analytical shock wave expression proves challenging. However, when the shock wave configuration depends solely on the $\tilde\psi$ coordinate -- corresponding to OTOC operators constrained to a Hopf circle with identical $(\theta,\phi)$ coordinates but separated exclusively along $\tilde\psi$ -- the calculation of~\eqref{eq:app:Hsol} simplifies dramatically. 
For simplicity, we may employ the isometry to set $\phi = \theta = 0$. For such configurations, in the large black hole limit ($r_h\gg L$), the shock wave profile becomes highly localized near the equatorial plane with $\theta\approx\theta_1 \approx 0$. 

Recalling that $d^{\mathcal{J}}_{\mathcal{KM}}(0)=\delta_{\mathcal{KM}}$, thus the infinite sum~\eqref{eq:app:Hsol} reduces to
\begin{align}
H(\alpha) =  E_0 e^{ 2\pi T t} \sum_{\mathcal{J}=0,1/2,1,\cdots}^{\infty} \sum_{\mathcal{K}=-\mathcal{J}}^{\mathcal{J}} \frac{2\mathcal{J}+1}{16\pi^2} \frac{e^{i\mathcal{K}\alpha}}{\mathcal{J}(\mathcal{J}+1)-\lambda_1^{sw} -i \lambda_2^{sw} \mathcal{K} + \lambda_3^{sw} \mathcal{K}^2} \,,
\end{align}
with $\alpha=\tilde\psi+\phi$. In the large black hole limit ($r_h \gg L$), where $\mathcal{J} \sim \mathcal{K} \sim r_h/L \gg 1$ with $\mathcal{K}\alpha$ held fixed, the above summation can be approximated by the integral given by
\begin{align}
 H(\alpha) =  E_0 e^{-2\pi T t} \int_0^{\infty} d\mathcal{J} \int_{-\mathcal{J}}^\mathcal{J} d\mathcal{K}  \frac{\mathcal{J}}{4\pi^2}  \frac{e^{i\mathcal{K}\alpha}}{\mathcal{J}^2-\lambda_1^{sw} -i \lambda_2^{sw} \mathcal{K} + \lambda_3^{sw} \mathcal{K}^2}  \,.
\end{align}
Considering the parameterization $\mathcal{K} = \mathcal{J}\cos\gamma$ with $\gamma\in [0,\pi]$, the integral takes the form
\begin{eqnarray}
\begin{aligned} \label{eq:app:Hinte}
 H(\alpha) &=  E_0 e^{2\pi T t} \int_0^{\infty} d\mathcal{J} \int_0^\pi d\gamma \frac{\mathcal{J}^2 \sin \gamma}{4\pi^2} \frac{e^{i\alpha \mathcal{J} \cos \gamma}}{\mathcal{J}^2-\lambda_1^{sw} -i \lambda_2^{sw} \mathcal{J} \cos \gamma + \lambda_3^{sw} \mathcal{J}^2 \cos^2 \gamma} \,, \\
 &= E_0 e^{-2\pi T t_w}  \int_0^{\pi/2} d\gamma \int_{-\infty}^{\infty} d\mathcal{J}  \frac{\mathcal{J}^2 \sin \gamma }{4\pi^2} \frac{e^{i\alpha \mathcal{J} \cos \gamma }}{\mathcal{J}^2-\lambda_1^{sw} -i \lambda_2^{sw} \mathcal{J} \cos \gamma + \lambda_3^{sw} \mathcal{J}^2 \cos^2\gamma }\,.
\end{aligned}
\end{eqnarray}
Note that we have changed the integration sequence, thereby ensuring the parameter $\cos\gamma$ of the exponent remains positive within the domain of integration.

Now we can compute $H$ using the contour integration and extract the quantum chaos parameters \emph{i.e.} the Lyapunov exponent $\lambda_L$ and butterfly velocity $v_B$. To compute~\eqref{eq:app:Hinte}, we will first integrate the $\mathcal{J}$ and subsequently evaluate the integral over $\gamma$. It is obvious that the integrand in~\eqref{eq:app:Hinte} has poles at $\mathcal{J}_{\pm}(\gamma)$ with
\begin{align}
    \mathcal{J}_{\pm}(\gamma) = \frac{i\lambda_2^{sw} \cos \gamma \pm \sqrt{4\lambda_1^{sw} -\left( \left(\lambda_2^{sw} \right)^2 -4\lambda_1^{sw} \lambda_3^{sw} \right) \cos^2\gamma} }{2 \left( \lambda_3^{sw} \cos^2 \gamma+1 \right)}\,.
\end{align}
The choice of integration contour depends crucially on the parameter $\alpha$. For $\alpha>0$, we consider a closed contour consisting of the real $\mathcal{J}$-axis $I_R$ and a semicircle in the upper half plane $I_C$. From the residue theorem, the resulting integral $I=I_R+I_C$ is given by
\begin{align}
\begin{split}
  I =I_R+I_C  &= - \frac{i E_0 e^{ 2\pi T t }  }{2\pi}    \int_0^{\pi/2} d\gamma  \frac{\mathcal{J}_+(\gamma)^2 \sin \gamma }{(1+\lambda_3^{sw}\cos^2 \gamma )\left( \mathcal{J}_+(\gamma )-\mathcal{J}_-(\gamma ) \right) } e^{i\alpha \mathcal{J}_+(\gamma) \cos t} \,,\\
  &= - \frac{i  E_0 e^{ 2\pi T t}  }{2\pi} \int_0^{\pi/2} d\gamma   \left( -\frac{1}{2} \frac{d \left( \mathcal{J}_+(\gamma ) \cos \gamma \right) }{d\gamma } \right)  e^{i\alpha \mathcal{J}_+(\gamma ) \cos \gamma} \,,\\
  &= \frac{ E_0 e^{ 2\pi T t }  }{4\pi \alpha} \left( 1 - e^{i\alpha \mathcal{J}_+(0)} \right)\,, \\
  &= \frac{  E_0 e^{ 2\pi T t }  }{4\pi \alpha} \left( 1 - e^{-\alpha  \mathcal{K}_{+,sw}} \right) \,,
\end{split}
\end{align}
where
\begin{equation}
\mathcal{K}_{+,sw} = -i \mathcal{J}_+(0) = \frac{\lambda_2^{sw} - \sqrt{ \left( \lambda_2^{sw} \right)^2 - 4\lambda_1^{sw} \left( 1+ \lambda_3^{sw} \right)}}{2 \left( \lambda_3^{sw}+1 \right)} \,.
\end{equation}
Recalling that the shock wave profile corresponds to $H(\alpha)=  \lim_{R\to\infty} I_R$. After a straightforward calculation of $I_C$ for $R\to\infty$, we can obtain the expression for the shock wave profile $H(\alpha)$. A similar computation of $H(\alpha)$ can be performed in the case of $\alpha<0$.

Finally, in the large black hole limit, the shock wave profile~\eqref{eq:app:Hsol} takes the following form
\begin{align}
 H(\alpha)=  \frac{  E_0 e^{ 2\pi T t }  }{4\pi \alpha} \bigg[  e^{\alpha k_-^{sw}} \Theta(-\alpha) - e^{-\alpha k_+^{sw}} \Theta(\alpha) \bigg] \,,
\end{align}
with $k_\pm^{sw} =\mathcal{K}_{\pm,sw}= -i \mathcal{J}_\pm(0)$.

\section{The CLP solution}\label{app:clp}
In this section, we provide details of the CLP solution. At the supergravity value $k=k_{susy}=2/\sqrt{3}$, the EMCS theory has an analytical charged rotating solution with equal angular momentum $J_1=J_2=J$, known as the CLP black hole~\cite{Cvetic:2004hs}. Expressed using the ansatz
\begin{eqnarray} 
\begin{split}
ds^2&=-F_0dt^2 +\frac{dr^2}{F_1} +\frac{F_2}{4} \left(\sigma_1^2+\sigma_2^2\right) +\frac{F_3}{4}\left(\sigma_3- \Omega dt \right)^2 \,,  \quad
A=A_t dt +\frac{1}{2} A_\psi \sigma_3 \,, 
\end{split} 
\end{eqnarray}
the CLP solution is given by~\cite{Madden:2004ym,Blazquez-Salcedo:2016rkj}
\begin{equation}
\begin{split}
F_1(r) &= 1-\frac{2m(1-\frac{a^2}{L^2})-2q}{r^2}+\frac{2a^2m+(1-\frac{a^2}{L^2})q^2}{r^4}+\frac{r^2}{L^2}, \\
F_2(r) &=r^2\,, \qquad F_3(r) = r^2({1+\frac{2a^2m}{r^4}}-\frac{a^2q^2}{r^6}) \,,\qquad 
F_0(r) =\frac{F_1(r)} {1+\frac{2a^2m}{r^4} -\frac{a^2q^2}{r^6}} \,,  \\
\Omega(r) &= \frac{ 2a(2m-q-\frac{q^2}{r^2})}{r^4 (1+\frac{2a^2m}{r^4}-\frac{a^2q^2}{r^6})} \,, \qquad A_t(r) = -\frac{\sqrt{3}q}{r^2} +\xi \,,\qquad  A_\psi(r)= \frac{\sqrt{3}aq} {r^2} \,,
\end{split}
\end{equation}
where $m, a$ and $q$ are parameters related to the mass, rotation and electrical charge of the solution.  The parameter $\xi$ in the component $A_t$ is introduced to fix the $U(1)$ gauge freedom as discussed in the main text, see~\eqref{eq:CLPgauge}. The AdS boundary is located at $r\to\infty$. There are several interesting limiting cases for CLP black holes. For $q=0$, the CLP black hole reduce to the rotating Myers-Perry-AdS$_5$ black holes with equal angular momentum~\cite{Hawking:1998kw}. Setting $a=0$, one obtain the electrically charged Reissner-Nordstr\"{o}em AdS$_5$ black holes.

In terms of the event horizon $r_h$ where $F_0(r_h)=F_1(r_h)=0$, the mass parameter $m$ can be expressed as
\begin{equation} \label{app:eq:mrh}
m= \frac{r_h^6 -a^2q^2 +L^2\left( q+r_h^2 \right)^2 }{ 2L^2r_h^2 -2a^2 \left(L^2 +r_h^2 \right) } \,,
\end{equation}
and the horizon angular velocity reads
\begin{align}
    \Omega_H  = \frac{2a(r_h^4+L^2(q+r_h^2))}{L^2(a^2q+r_h^4)}\,.
\end{align}
The temperature associated with the Killing vector $\chi = \partial_t + \Omega_H \partial_{\psi}$ is
\begin{align}
2\pi T &= \frac{ 2L^2 r_h^6 -a^4 q^2 - L^4(q^2-r_h^4) - 2a^2 \left(r_h^6+L^4(q+r_h^2) +L^2(q^2-2 r_h^4) \right) }{L^3(a^2q+r_h^4) \sqrt{L^2r_h^2 - a^2(L^2+r_h^2)}}\,, 
\end{align}
and the entropy reads
\begin{align}
S = \frac{A_H}{4 G_N} = \int  \frac{F_2\sqrt{F_3} \sin\theta }{32 G_N} d\theta d\phi d\psi   \bigg|_{r=r_h}=  \frac{\pi^2 L (a^2q+r_h^4)}{2G_N \sqrt{L^2r_h^2 - a^2(L^2+r_h^2)}}\,.
\end{align}
It is worth noting that the parameters $(r_h, a, q)$ must satisfy the constraint
\begin{align}
a^4 q^2 - 2L^2 r_h^6 + L^4(q^2-r_h^4)+ 2a^2 \left(r_h^6+L^4(q+r_h^2) -L^2(q^2-2 r_h^4) \right) \leq 0\,,
\end{align}
such that $T\ge 0$.
When the above inequality saturates, the black hole horizon is degenerate and the zero temperature extremal black hole is realized. The electrostatic potential $\Phi$, as measured at infinity with respect to the horizon, reads
\begin{align}
\Phi = \chi^bA_b \big|_{r\to\infty} - \chi^bA_b \big|_{r=r_h} = \frac{\sqrt{3}q \left( L^2r_h^2 -a^2(L^2+r_h^2) \right)}{  L^2 ( a^2 q +r_h^4) } \,.
\end{align}

Other conserved charges of the rotating black hole can be obtained using the holographic renormalization~\cite{Balasubramanian:1999re,deHaro:2000vlm}. Note that, the holographic renormalization approach for thermodynamics was also applied to rotating black hole with three independent electric charges in~\cite{Cassani:2019mms}, which reduces to the CLP solution upon setting the electric charges equal. After adding additional surfaces terms to the original action, the renormalised on shell action is given by
\begin{equation}\label{app:eq:sren}
S_{ren}= S +S_{bdy}\,,
\end{equation}
where 
\begin{equation} \label{app:eq:sbulk}
S= \frac{1}{16\pi G_N} \int d^5x \sqrt{-g} \Big( R+\frac{12}{L^2}- \frac{1}{4} F_{ab}F^{ab}  +\frac{k}{24} \epsilon^{abcde}A_a F_{bc} F_{de} \Big)\,,
\end{equation}
and
\begin{equation} \label{app:eq:sbdy}
S_{bdy}=\frac{1}{16\pi G_N}\int d^4x\sqrt{-\gamma} \left( 2K -\frac{6}{L} -\frac{L\hat{R}}{2} \right)\,,
\end{equation}
with $\gamma_{\mu\nu}$ the induced metric at the AdS boundary, $K$ the trace of extrinsic curvature $K_{\mu\nu}$ and $\hat{R}_{\mu\nu}$ the Ricci tensor associated with $\gamma_{\mu\nu}$. Using the properties of background equations~\eqref{eq:eom} as well as applying the identity $R^a_{\ b}\xi^b=\nabla_b\nabla^a\xi^b$ on a timelike killing vector $\xi^a=(\partial_t)^a$, the bulk on-shell action takes
\begin{equation} \label{app:eq:sbulkos}
S= \frac{1}{16\pi G_N} \int d^5x \frac{\sin\theta}{8} \left[\sqrt{F_2 } F_3 \left[ \frac{F_3\Omega \Omega'}{4} -F_0' +\frac{A_t}{2} \left( 2A_t' +\Omega A_\psi'\right) \right] -\frac{2 k A_t A_\psi^2}{3} \right]' .
\end{equation}
The stress tensor and current of the dual field theory are
\begin{equation}
\begin{split}
\langle T_{\mu\nu} \rangle &= \frac{1}{16\pi G_N} \lim_{r\to\infty} \frac{r^2}{L^2} \bigg[ -2 (K_{\mu\nu} -K \gamma_{\mu\nu}) -\frac{6}{L} K\gamma_{\mu\nu} +L\hat{G}_{\mu\nu}  \bigg] \,, \\
\langle J^{\mu} \rangle  &= \frac{1}{16\pi G_N} \lim_{r\to\infty} \frac{r^4}{L^4} \Big[ n_r \Big( F^{\mu r} +\frac{k}{6}\epsilon^{r\mu \alpha\beta\gamma} A_\alpha F_{\beta\gamma} \Big) \Big] \,.
\end{split}
\end{equation}
The non-zero components of the stress tensor and current  are
\begin{equation}
\begin{split}
& \epsilon=  \langle T_{tt}\rangle= \frac{1}{16\pi G_N} \left[ \frac{6(m-q) }{L^3 } +\frac{2m a^2 }{L^5 } +\frac{3}{4L} \right] \,,\\ 
& \langle T_{\theta\theta}\rangle = \frac{1} {16\pi G_N} \left[\frac{m-q}{2L} -\frac{m a^2}{2L^3} +\frac{L}{16} \right] \,, \\
& \langle T_{\phi\phi}\rangle = \frac{1}{16\pi G_N} \left[ \frac{m-q}{2L} +\frac{m a^2(1+ 2\cos{(2\theta)}) }{2L^3 } +\frac{L}{16} \right] \,, \\
& \langle T_{\psi\psi}\rangle= \frac{1}{16\pi G_N} \left[\frac{m-q}{2L} +\frac{3m a^2}{2L^3} +\frac{L}{16} \right] \,, \\
& j=\langle T_{t\psi}\rangle= \langle T_{\psi t}\rangle= \frac{}{} -\frac{a(2m-q) }{ 8\pi G_N L^3} \,, \\
& \langle T_{t\phi}\rangle=\langle T_{\phi t}\rangle= \cos\theta \langle T_{t\psi}\rangle \,, \qquad  \langle T_{\phi\psi}\rangle=\langle T_{\psi\phi}\rangle= \cos\theta \langle T_{\psi\psi}\rangle   \,, \\
& \langle J^t \rangle = \frac{ \sqrt{3} q }{ 8\pi G_N L^3} \,, \quad
\langle J^\psi \rangle = \frac{ \sqrt{3} a q }{ 4\pi G_N L^5}  \,. 
\end{split}
\end{equation}
The conserved charge associated with the isometry generated by a Killing vector $\xi$ on the boundary geometry is defined as~\cite{Balasubramanian:1999re}
\begin{equation}
Q_\xi = \int_{S^3_\infty} u^a\xi^b \langle T_{ab}\rangle d^3S \,,
\end{equation}
where $u_a=\delta^t_a$ is a timelike unit normal vector. The mass is the conserved charge for $\xi_{(t)}^a=(\partial_t)^a$, while the angular momentum is associated with $\xi_{(\psi)}^a=(\partial_\psi)^a$. A similar definition holds for the conserved electric charge $Q$ of the $U(1)$ current.
Therefore, the total mass, angular momentum and electric charge of the CLP black hole are given by
\begin{eqnarray} \label{app:eq:MJQ}
\begin{split}
M  &=  \int_{S^3_\infty} u^a \xi_{(t)}^b \langle T_{ab} \rangle d^3S = V \epsilon = \frac{\pi}{4G_N} \left( 3\left( m-q \right) +\frac{ma^2}{L^2} +\frac{3L^2}{8}\right) \,, \\
J &=  \int_{S^3_\infty} u^a \xi_{(\psi)}^b \langle T_{ab} \rangle d^3S= -V j =  \frac{\pi}{4G_N} a \left(2m-q\right) \,,\\
Q &= \int_{S^3_\infty} u_a \langle J^a\rangle d^3S = V \langle J^t\rangle = \frac{ \sqrt{3}\pi }{4G_N} q \,,
\end{split}
\end{eqnarray}
where the spatial volume of the boundary metric is $V= \frac{1}{8}\int \sin\theta L^3 d\theta d\phi d\psi =2\pi^2 L^3$. The expression for the mass presented here differs from that derived in~\cite{Madden:2004ym} by a constant term $\frac{3\pi L^2}{32G_N}$. This term corresponds to the energy of global $AdS_5$ spacetime and represents the Casimir energy of the dual conformal field theory defined on $R\times S^3$~\cite{Balasubramanian:1999re,Gutowski:2004ez}.

Furthermore, one can find that the thermodynamic quantities presented above satisfy the first law of thermodynamics for charged rotating black holes~\cite{Madden:2004ym}, \emph{i.e.}
\begin{eqnarray} \label{app:eq:firstlaw}
\begin{split}
dM &= TdS +\Phi dQ +\Omega_H dJ \,.
\end{split}
\end{eqnarray}
A convenient way to verify~\eqref{app:eq:firstlaw} is to use~\eqref{app:eq:mrh} to eliminate $m$ in~\eqref{app:eq:MJQ} and then to consider the variation of thermodynamic variables $(M,S,Q,J)$ with respect to $(a,q,r_h)$.

Finally, performing a Wick rotation $t=-i\tau_E$ and $a=-ia_E$, we obtain the Euclidean action
\begin{equation}
\begin{split}
I_E & = I+I_{bdy}  \\
&= \frac{ \pi  \Delta\tau_E }{32G_N} \bigg[ 8\left(m-q -\frac{ma^2}{L^2}\right) +3L^2 +\frac{8a^2q^2}{r_h^2 L^2} -\frac{8r_h^4}{L^2} \\
&\quad\quad - \frac{8q^2 \left( r_h^6 -2ma^4 +a^2(2m-q)r_h^2 -a^2(q^2+r_h^4) \right)}{ r_h^8 +a^2r_h^2 (2mr_h^2 -q^2)} \bigg] \,,
\end{split}
\end{equation}
where $I=-iS$ and $I_{bdy}$ denotes the Euclidean boundary action derived from~\eqref{app:eq:sbdy}. The period of the coordinate $\tau_E$ is $\Delta\tau_E=\int d\tau_E =T^{-1}$. In the grand canonical ensemble, the corresponding free energy $W$ is given by
\begin{equation}\label{eq:free}
W= -T \ln Z=  T I_E  \,.
\end{equation}
Using the thermodynamic variables we have defined above, we obtain the standard thermodynamic relation
\begin{equation}
W = M -TS -\Phi Q -\Omega_H J \,,
\end{equation}
together with the expected form of the  first law of thermodynamics
\begin{equation}\label{firstlaw}
dW= -S dT -Q d\Phi -J d\Omega_H \,.
\end{equation}
It is worthy to note that the introduction of the $U(1)$ gauge $\xi$ does modify the on-shell action, as can be directly observed from~\eqref{app:eq:sren},~\eqref{app:eq:sbulk} and~\eqref{app:eq:sbulkos}. Moreover, such gauge fixing is crucial for evaluating the on-shell Euclidean action and obtaining the correct free energy~\eqref{eq:free}. In contrast, if one instead adopts a different gauge--for instance, setting $\xi = 0$--the resulting free energy would become inconsistent with the first law of thermodynamics~\eqref{firstlaw}. This further reinforces the validity of the boundary condition motivated by the Wilson loop argument presented in Section~\ref{sec:5}.

\bibliographystyle{JHEP}
\bibliography{crBH}
\end{document}